\documentclass[acmsmall, screen]{acmart}

\AtBeginDocument{%
  \providecommand\BibTeX{{%
    \normalfont B\kern-0.5em{\scshape i\kern-0.25em b}\kern-0.8em\TeX}}}

\setcopyright{none}
\renewcommand\footnotetextcopyrightpermission[1]{} 

\usepackage{scalerel}
\usepackage{multirow}
\usepackage{enumitem}
\usepackage{listings}
\usepackage{xparse}
\usepackage[utf8]{inputenc}
\usepackage{xcolor}
\usepackage{colortbl}
\usepackage{makecell}
\usepackage[nameinlink]{cleveref}
\usepackage{array}
\usepackage{wrapfig}
\usepackage{lipsum}
\usepackage{comment}
\usepackage[normalem]{ulem}
\usepackage{hyperref}
\usepackage{hyperxmp}
\usepackage{subcaption}
        
\begin{document}

\title[RAI Practices in Industry]{What We Know about Responsible AI Practices in Industry: \\A Half Decade of Empirical Research}
%


\author{Wesley Hanwen Deng}
\orcid{0000-0003-3375-5285}
\email{wesleyhdeng@gmail.com}
\affiliation{
\institution{Microsoft Research}
  \streetaddress{300 Lafayette St.}
  \city{New York}
  \state{NY}
  \postcode{10012}
  \country{USA}
}

\author{Agathe Balayn}
\orcid{0000-0003-2725-5305}
\email{balaynagathe@microsoft.com}
\affiliation{
\institution{Microsoft Research}
  \streetaddress{300 Lafayette St.}
  \city{New York}
  \state{NY}
  \postcode{10012}
  \country{USA}
}

\author{Andrew Selbst}
\orcid{} 
\email{aselbst@law.ucla.edu}
\affiliation{
\institution{University of California, Los Angeles}
  \streetaddress{405 Hilgard Ave.}
  \city{Los Angeles}
  \state{CA}
  \postcode{90095}
  \country{USA}
}

\author{Jason I. Hong}
\orcid{0000-0002-9856-9654}
\email{jasonh@cs.cmu.edu}
\affiliation{%
  \institution{Carnegie Mellon University}
  \city{Pittsburgh}
  \state{Pennsylvania}
  \country{USA}}

\author{Motahhare Eslami}
\orcid{0000-0002-1499-3045}
\email{meslami@cs.cmu.edu}
\affiliation{%
  \institution{Carnegie Mellon University}
  \streetaddress{5000 Forbes Ave}
  \city{Pittsburgh}
  \state{PA}
  \postcode{15213}
  \country{USA}
}

\author{Kenneth Holstein}
\orcid{0000-0001-6730-922X}
\email{kjholste@cs.cmu.edu}
\affiliation{%
  \institution{Carnegie Mellon University}
  \streetaddress{5000 Forbes Ave}
  \city{Pittsburgh}
  \state{PA}
  \postcode{15213}
  \country{USA}
}

\author{Hanna Wallach}
\orcid{} 
\email{wallach@microsoft.com}
\affiliation{
\institution{Microsoft Research}
  \streetaddress{300 Lafayette St.}
  \city{New York}
  \state{NY}
  \postcode{10012}
  \country{USA}
}

\author{Jennifer Wortman Vaughan}
\orcid{0000-0002-7807-2018}
\email{jenn@microsoft.com}
\affiliation{
\institution{Microsoft Research}
  \streetaddress{300 Lafayette St.}
  \city{New York}
  \state{NY}
  \postcode{10012}
  \country{USA}
}

\author{Solon Barocas}
\orcid{0000-0003-4577-466X}
\email{solon@microsoft.com}
\affiliation{
\institution{Microsoft Research}
  \streetaddress{300 Lafayette St.}
  \city{New York}
  \state{NY}
  \postcode{10012}
  \country{USA}
}

\renewcommand{\shortauthors}{Deng et al.}

\begin{abstract}
  Responsible AI (RAI) has become a central concern for technology companies, regulators, and the public. How industry practitioners interpret, implement, and sustain RAI work directly shapes the design and deployment of AI systems. As empirical scholarship examining RAI practices in industry has rapidly expanded, findings are dispersed across studies that focus on different roles, organizational contexts, and interventions. This work synthesizes current knowledge through a literature review of 161 empirical studies spanning six years, each engaging industry practitioners via interviews, surveys, workshops, ethnographies, and other methods.
  Our synthesis reveals both meaningful progress and persistent challenges in industry RAI practice. Practitioner awareness has increased, RAI activities have become more professionalized, and interventions such as toolkits and guidelines are more widely adopted. At the same time, practitioners continue to face substantial barriers, including limited training, uneven organizational support, and a lack of interventions tailored to day-to-day work practices. By consolidating and organizing these findings, we provide a more complete account of industry RAI than any single study to date. We conclude by discussing implications for RAI researchers, practitioners seeking to adopt effective practices, and policymakers aiming to ground governance efforts in the realities of industry contexts.
  
\end{abstract}



\maketitle

\section{Introduction}

The safe and responsible design, development, and deployment of AI systems has become a growing priority across industry, policy, and public discourse. In practice, responsible AI (RAI) is enacted not through abstract principles alone, but through the everyday decisions, trade-offs, and workflows of industry AI practitioners. Understanding how RAI is operationalized in organizational settings is therefore essential to understanding how AI systems are actually built, governed, and experienced in the world. \looseness=-1

In response to this need, empirical research examining RAI practices in industry has expanded rapidly in recent years. This body of work spans a wide range of practitioner roles, organizational contexts, and methodological approaches, offering valuable but often partial views into RAI as it is practiced. As a result, insights about practitioner motivations, challenges, tool use, and organizational dynamics remain distributed across individual studies, making it difficult to form a coherent picture of how RAI practices function and evolve over time. What is currently missing is a consolidated account that takes stock of what this growing empirical literature collectively reveals about RAI in industry.

Synthesizing prior work about how RAI is carried out in industry and building a landscape of RAI practice from the existing literature is critical for several reasons. For \textbf{practitioners}, a synthesized understanding can surface shared challenges, emerging norms, and promising approaches that may inform their own RAI efforts. For \textbf{researchers}, this work can help identify gaps where current tools, frameworks, and methods fail to align with real-world needs, guiding the design of more actionable interventions. Such synthesized research landscape is also essential for researchers to identify research gaps, surface methodological and conceptual trends, and carry out future inquiry in this rapidly evolving area. For \textbf{policymakers}, it can ground governance and regulatory proposals in the realities of industry practice, helping to avoid implementing policies that are either too high-level or disconnected from day-to-day implementation.

To contribute to this understanding, in this work, we explore the following two research questions:
\begin{itemize}[leftmargin=*]
    \item What does prior literature reveal about the current practices and challenges that industry practitioners face when doing RAI on the ground?
    \item Based on this literature, how might future researchers, practitioners, and policymakers better support RAI in industry practice?
\end{itemize}

To address these questions, we conducted a literature review 
of 161 empirical studies on industry RAI across research communities Between 2019 and August 2025. The authors read and analyzed all 161 papers without any support from generative AI.
We define empirical work as research involving industry practitioners through methods such as interviews, surveys, workshops, ethnographic studies, or other standard empirical approaches. Our search aimed to encompass a broad range of RAI topics, including AI fairness, transparency, explainability, interpretability, accountability, privacy and security, accessibility, ethics, safety, and more. After building the full research corpus, we annotated each paper for both descriptive data, such as publication details and information about research participants and methods, as well as the main findings of these works, including the concrete empirical findings and implications reported. 

Based on the empirical findings of these papers, we identified current practices that demonstrate progress towards improved RAI practices in industry, while also highlighting persistent challenges tied to these practices. In particular, we find that while practitioners' awareness regarding the existence and importance of RAI has grown (Section \ref{practice:awareness}) and RAI activities have become more professionalized and routine (Section \ref{practice:professionalized}), practitioners often still lack the expertise, training, and organizational support needed to carry out this work effectively (Sections \ref{challenge:lack of expertise} and \ref{challenge:organizational dynamics}). Furthermore, while many RAI interventions—such as computational toolkits, guidelines, documentation, and policies—are increasingly adopted in industry (Section \ref{practice:RAI-interventions}), there remains a gap in tailoring these interventions to specific AI applications (the use cases or products where AI is deployed), domains (the fields or industries those applications belong to), and pipelines (the workflows and stages in model design, development, and deployment) (Section \ref{challenge:tailored interventions}). Finally, although practitioners are beginning to engage with external stakeholders such as data annotators, end users, and domain experts (Section \ref{practice:external}), these efforts are frequently marked by frictions for meaningfully engaging  (Section \ref{challenge:external friction}). \looseness=-1

 \begin{figure*}[t]
  \centering
  \includegraphics[width=1\linewidth]{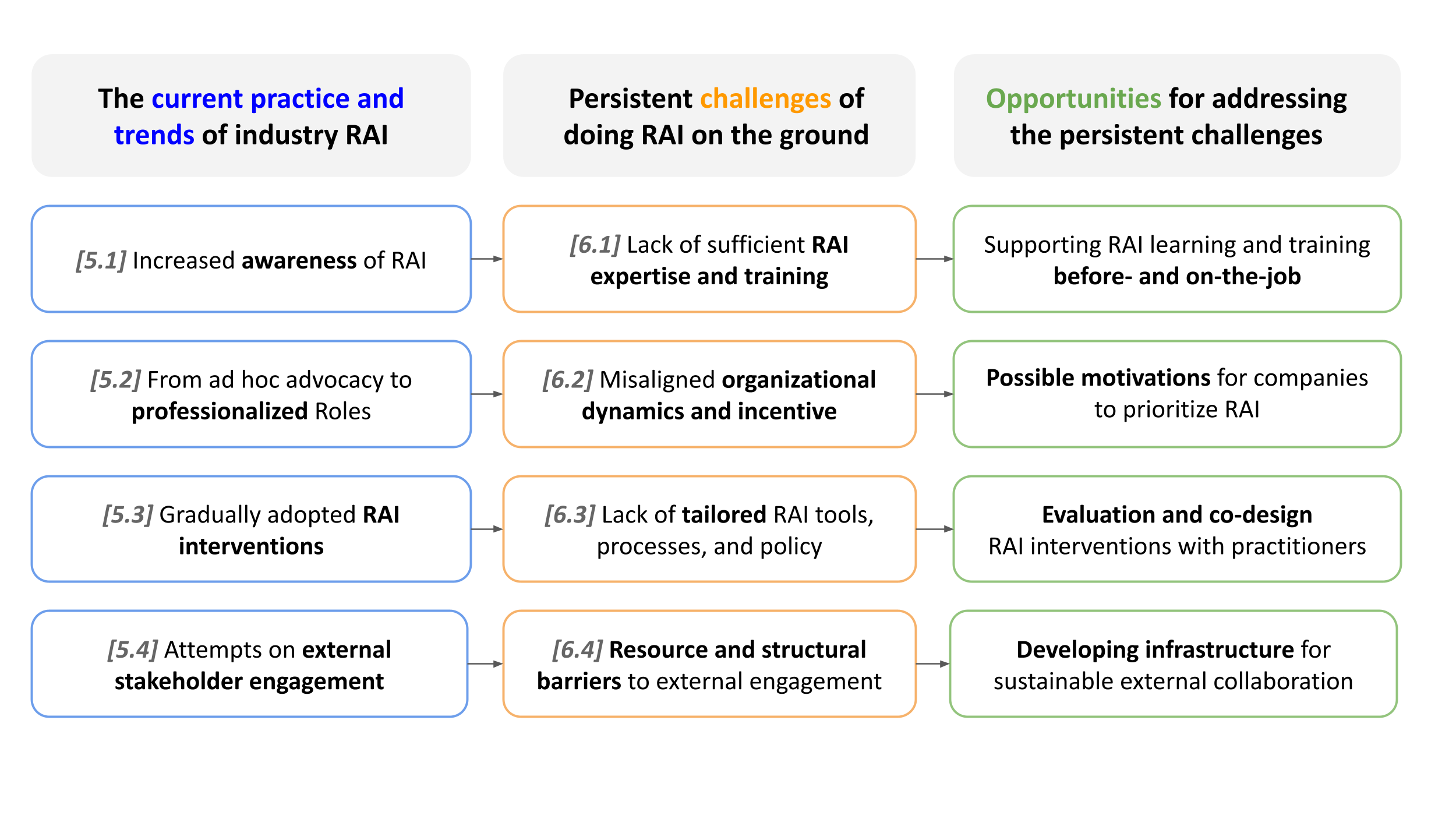}
  \caption{
    An overview of the main findings of our review of 161 
    empirical works on industry RAI practice. We identify current practices that demonstrate progress in improving RAI in industry, while also highlighting persistent challenges tied to these practices. On the left side, we present the synthesized findings on current industry RAI practices (Section \ref{practice}). On the right side, we present the challenges corresponding to these ``success stories'' of doing RAI in industry settings, as well as the key opportunities suggested by the papers we reviewed (Section \ref{challenge}). 
}
  \Description{TBA }
  \label{fig:result overview}
\end{figure*}

For each challenge, we synthesize corresponding opportunities researchers have proposed in response, highlighting both areas of agreement and disagreement on potential solutions (Section \ref{challenge}). Building on these findings, we discuss concrete implications for researchers, practitioners, and policymakers seeking to better support RAI in industry practice (Section \ref{discussion}). These include call for moving beyond cataloging known problems toward designing, deploying, and evaluating RAI interventions as end-to-end sociotechnical systems embedded in real organizational workflows — rather than isolated tools or compliance artifacts primarily focusing on assessing model outputs. In addition, sustaining this shift requires not only methodological investment in longitudinal and ethnographic approaches from researchers, but also regulatory frameworks that prioritize implementability and substantive accountability over procedural compliance, and that leverage the signaling power of anticipated regulation to shape organizational RAI capacity-building. \looseness=-1

Our paper ultimately makes the following contributions:

\begin{itemize}[leftmargin=*]
    
    \item \textbf{An in-depth synthesis of knowledge of the state of RAI in industry practice}, drawing on 161 prior empirical studies of industry practitioners across diverse research communities. In particular, we systematically reviewed, annotated, and analyzed this body of work to surface concrete practices, persistent challenges, and opportunities for implementing RAI on the ground.

    \item \textbf{A set of actionable implications and recommendations} tailored for industry practitioners, RAI researchers, and policymakers. These insights highlight opportunities for different actors to improve the status quo in RAI practice, RAI research, and AI regulation. 

    \item \textbf{An open-source database}\footnote{Link to database redacted to maintain anonymity during review.} of all 161 existing empirical research papers on RAI in industry practice we reviewed in this work. This resource is designed to support RAI researchers, practitioners, and policymakers in identifying relevant studies, avoiding duplication of prior work, and building on cumulative knowledge. 
\end{itemize}

\section{Background} \label{sec:rw}

A number of existing surveys tackle questions adjacent to ours. To start with, a number of publications have reviewed how RAI and adjacent terms such as AI ethics and Trustworthy AI  
have been defined \cite{goellner2024responsible,gunasekara2025systematic} and specified into RAI principles \cite{correa2023worldwide,jobin2019global}, what RAI tools (including user interfaces and governance frameworks) \cite{ortega2024applying,lu2024responsible,zhou2020survey}, guidelines \cite{jobin2019global,morley2020initial,al2023scoping}, and technical approaches to mitigate RAI-related issues (e.g., \cite{palumbo2024objective,zhang2024ai}) have been developed, and to what extent these tools and approaches cover existing lists of RAI principles \cite{prem2023ethical}. Surveys also discuss \emph{analytically} what challenges might arise when operationalizing these principles in practice \cite{woodgate2024macro}. However, these works do not \emph{empirically} examine the current practices of the industry practitioners who might adopt these tools or draw on such guidelines to meet the RAI principles set by their organizations. \textbf{Our work} instead focuses exclusively on publications reporting empirical work with AI practitioners, allowing us to map the state of RAI ``on the ground'' in terms of both current practices and the challenges practitioners face.

Some of these literature reviews have focused on specific sub-topics of RAI, such as explainable AI \cite{arrieta2020explainable} or AI fairness \cite{mehrabi2021survey}, while most others do not differentiate between these topics and include them all together under umbrella terms like “responsible AI,” “trustworthy AI,” or “ethical AI.” In contrast, \textbf{our work} deliberately takes a broad view of RAI, examining RAI concepts holistically rather than limiting our scope to individual values such as fairness or explainability. Doing so enables us to identify similarities and differences in the ways these notions are tackled in practice (e.g., similar organizational challenges), while making sure that no RAI notion falls in the cracks of our survey. 

Finally, a few recent publications \cite{khan2022ethics,pant2024ethics,anagnostou2022characteristics,bach2025insights,sadek2025challenges}
are more closely related to our study, as they might share similar goals and investigate a similar corpus to ours. However, \textbf{our work} differs in three key ways. First, we restrict our corpus to publications presenting empirical studies with AI practitioners, and we ensure coverage by searching across multiple databases and disciplines. This approach yields a more comprehensive yet focused set of publications. Second, these recent reviews typically organize results by broad RAI categories or use cases, without discussing the themes that emerge across papers and research communities.
In contrast, we identify key practices and challenges faced by AI practitioners across papers, tracking their evolution over time. We also assess the prevalence and severity of these challenges, distinguishing between types of practitioners and AI pipeline phases addressed. Finally, based on our analysis, we explicitly articulate actionable implications for researchers, policymakers, and practitioners, going beyond listing high-level challenges to highlight concrete opportunities for addressing them. Together, these contributions provide a more comprehensive and practice-grounded synthesis of empirical RAI research across different research communities than existing surveys.

\section{Methods} \label{sec:methods}

To investigate our research questions, we include
peer-reviewed research papers published between 2018 and August 2025 that employ empirical methods (e.g., interview studies, surveys, workshops) to provide insights into the current practices, challenges, and opportunities for industry practitioners in conducting responsible AI work on the ground. To develop a comprehensive corpus of empirical research on industry RAI practices, we followed a literature review process inspired by prior work \cite{stefanidi2023literature,mack2021we,gencc2024situating,lopez2017awareness}, as described below. 

\subsection{Collecting Publications} \label{collecting pub}

To develop our corpus of papers, we first generated an initial corpus using a mix of a manually assembled collection of  seed papers (a ``seed set'' of relevant papers already known to the authors), keyword searches in computer science databases, and targeted searches in additional databases. We describe each of these three strategies below. \looseness=-1

 \begin{figure*}[t]
  \centering
  \includegraphics[width=1\linewidth]{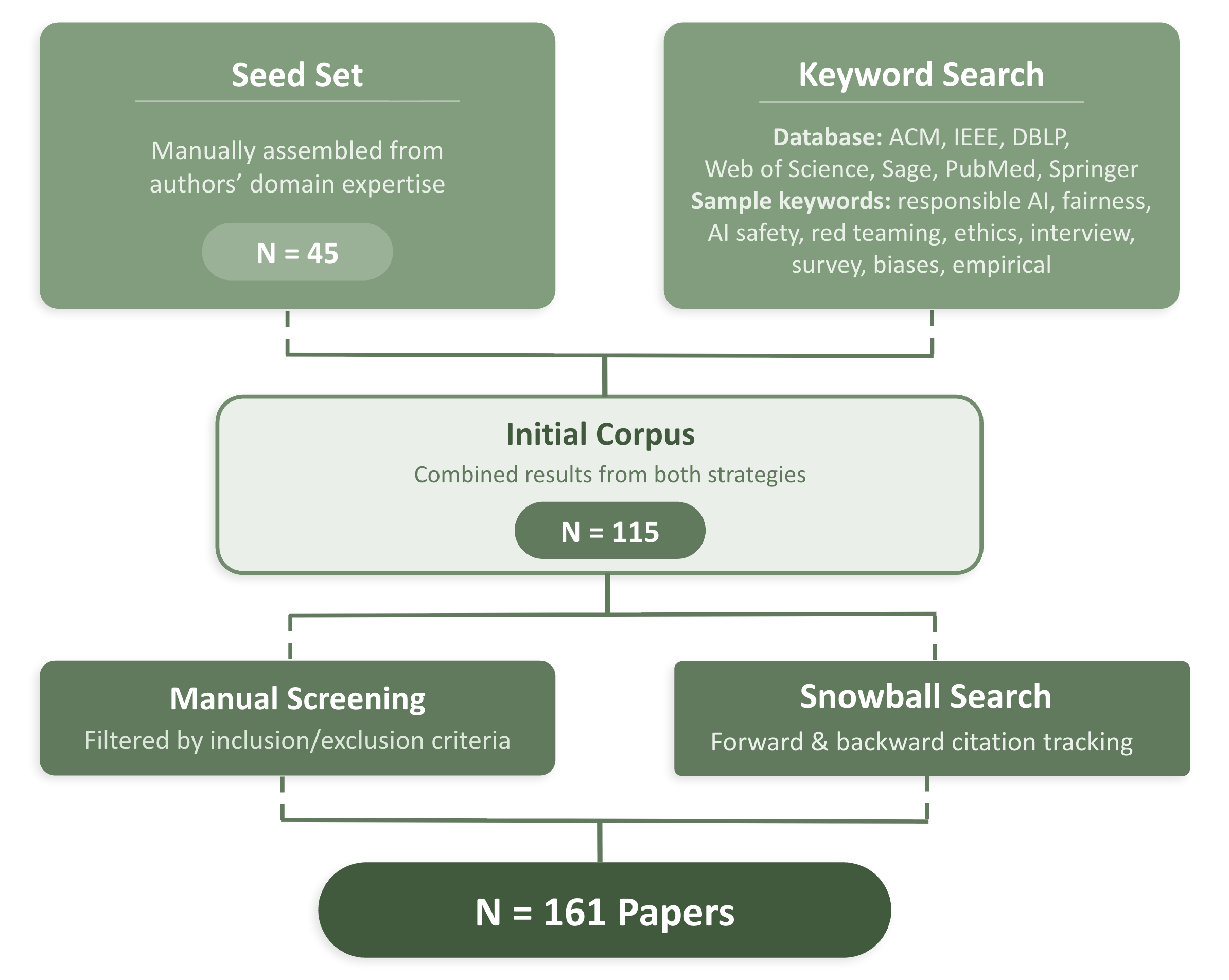}
  \caption{
    An overview of our paper review process
}
  \label{fig:result overview}
\end{figure*}

We manually created the seed set (N = 45) by collecting papers describing empirical studies of RAI practices in industry that met our criteria (e.g., \cite{holstein2019improving, madaio2020co, orr2020attributions, lee2021landscape}) based on our own knowledge of the space. (All authors have published papers that meet our criteria in the past.) While assembling this seed set, we began to iterate as a team on our concrete inclusion and exclusion criteria (see Section \ref{method:criteria}) and potential keywords to search. 

Following the traditional systemic literature review process~\cite{keele2007guidelines,macdonell2010reliable}, 
we then conducted keyword-based searches in the major computer science databases: ACM digital library, IEEE, DBLP, and Semantic Scholar. We queried these databases using keywords related to responsible AI—ensuring coverage of the diverse terminologies commonly used to describe RAI-relevant topics—as well as empirical research. Our queries varied slightly across databases to account for differences in supported query expressivity. For example, we used keywords such as “empirical,” “qualitative,” “quantitative,” “interview,” and “survey” to surface studies involving empirical research. We also included topic-specific keywords such as “responsible AI,” “RAI,” “fairness,” “harm,” “ethics,” “biases,” “safety,” and “red teaming” to identify relevant research areas. We include our full keyword strings in Appendix \ref{appendix}. \looseness=-1

Finally, as some papers about RAI are published in disciplines beyond HCI and AI, we conducted targeted searches in specialized databases that were likely to include relevant publications. Such databases included Web of Science, Sage, PubMed, and Springer. As many of these databases do not allow for inputting expressive queries, the results were too numerous to review exhaustively. As a result, we only analyzed the first 300 results for each query on the assumption that the more relevant results would appear earlier in the list, selected the relevant publications, and noted the relevant journals in which they were published. Then, we queried each journal individually (with much simpler queries), and collected all relevant publications.

With the initial corpus created by these three strategies, we then conducted a snowball search, and identified additional relevant papers cited by or citing these initial papers using Google Scholar. For all new papers identified through this process, we repeated the citation tracking process. We then manually screened all retrieved papers and filtered out irrelevant ones based on our inclusion criteria. This iterative process continued until no new relevant papers were found, ensuring comprehensive coverage of the available literature. Through this method, we collected in total 161 papers focusing on empirical studies of RAI practices in industry. 

We conducted this process between October 2024 and August 2025, and included publications from 2018 to 2025. We began our paper search in 2018, as this was the year when most technology companies started actively engaging in responsible AI.\footnote{The earliest papers we found are from 2019.} The cut-off date of August 2025 reflects the month before we transitioned from data collection to data analysis and paper writing.

\subsection{Refining Inclusion Criteria} \label{method:criteria}

Throughout the process of compiling the paper list, four authors met frequently to discuss and refine the inclusion and exclusion criteria. To address our research questions, we decided to include \textbf{peer-reviewed research papers published between 2018 and August 2025 that employ \textit{empirical methods} to study the \textit{practices of industry practitioners} engaged in \textit{responsible AI} work}. Our inclusion criteria emphasize three key points that helped to exclude several works that might otherwise have appeared relevant:

\begin{itemize}[leftmargin=*]
    \item \textbf{Empirical methods.} We focus on studies that employ empirical methods (e.g., interviews, workshops, surveys) to investigate practices. This excludes papers that only develop or evaluate tools or algorithms for responsible AI without engaging in empirical studies with human subjects (e.g., \cite{nguyen2024literature,ahmad2021human}), as well as papers offering case studies that do not involve empirical methods with human participants (e.g., \cite{quinonero2023disentangling,beutel2019putting}).
    \item \textbf{Industry settings.} We include only studies of industry contexts. Papers on responsible AI that examine practices and challenges faced solely by public sector developers, academic researchers, crowdworkers, or end users of AI systems are excluded unless they involve industry practitioners (e.g., \cite{adamyk2024barriers,das2023artificial}).
    \item \textbf{AI practices.} We focus specifically on AI practices. Studies that address concepts such as fairness, transparency, privacy, or safety but concentrate on general software engineering practices unrelated to AI or RAI are excluded (e.g., \cite{khanam2023understandings,charlesworth2004south}).
\end{itemize}

\subsection{Analyzing our Corpus}
To analyze the publications in our corpus, two authors first annotated them in a structured manner. For each publication, we noted major methodological choices, main and secondary results reported, and explicit discussions and recommendations made by the authors of the publication.
We then conducted a thematic analysis of the annotations in three steps.
\emph{1)} The two authors started clustering the annotations individually and then in collaboration to form initial themes based on the knowledge they acquired while annotating the publications.
\emph{2)} The other authors were introduced to the initial themes, and were allocated five publications to familiarize themselves with. 
Individually,  they were invited to record on post-its relevant annotations about each publication, and to position these post-its within the relevant themes if possible, and otherwise to reflect on potentially new clusters of post-its.
All the authors then discussed together this annotation and clustering exercise, and how the themes should be amended and what new themes should be added to better reflect the content of the corpus. 
\emph{3)} Based on these discussions, the first two authors then reviewed the remaining publications to enrich each theme and consolidate sub-themes.
In the next sections, we describe the themes and sub-themes that we landed on. Finally, we also calculated the descriptive statistics.

\section{Overview of Industry Empirical RAI Papers} \label{paper overview}

This section provides a descriptive analysis of the 161 papers identified in our review, mapping the growth, geographic distribution, and methodological trends within the field of empirical research focused on industry RAI practices. 

The temporal distribution of the literature indicates a significant and accelerating interest in industrial RAI. While the field was nascent in 2019 with only two identified papers, the volume of research grew steadily to 11 papers in 2020 and 21 in 2022. The most substantial surge occurred in the last two years, with 37 papers published in 2023 and 49 in 2024. As of August 2025, 21 papers have already been published. 

Regarding authorship and institutional representation, the majority of the reviewed papers (87\%) involve researchers from academia, while approximately 20\% feature contributions from researchers with industry affiliation. Geographically, the United States remains the primary hub for this research in terms of \textbf{author affiliations}, accounting for 45.3\% of the corpus (73/161 papers). The United Kingdom follows with 16.7\%, while Finland, Canada, Germany, and the Netherlands also emerged as significant contributors, each producing approximately a dozen papers. For participants demographic, around 80\% fo the paper (130/161 papers) include participants from the United States, roughly 38.5\% of the corpus (62/161 papers) included participants from the United Kingdom or Europe. Only 9 papers included participants from Asia. Together, these distributions suggest that while the discourse is global, both the production of RAI scholarship and the populations it studies remain concentrated within a few key Western technological hubs.

Methodologically, interviews were the primary instrument (n = 106), involving an average of 27 participants ($SD = 25.79$, range: 12–55). Surveys were utilized in 28 papers, featuring a mean of 201 subjects; however, the substantial standard deviation ($SD = 354$) reflects significant variance, as several studies included samples as large as 1,000 participants. Workshop-based research (n = 26) targeted smaller, more consistently sized cohorts, averaging 12 participants ($SD = 1.4$). Notably, papers often included studies with multiple research designs; for instance, 18 of the papers utilizing workshops also incorporated interviews as a primary or secondary method. Conversely, more specialized or resource-intensive methodologies were rare: only three papers reported using contextual inquiry, and only two conducted controlled experiments. This methodological landscape underscores a current research trend on gathering practitioner perspectives through direct dialogue rather than through observational or experimental studies.

\section{Current Practices and Trends in RAI in Industry} \label{practice}

In this section, we describe the positive trends identified across the corpus. These trends reflect how individual AI practitioners have developed a greater understanding of RAI over time (Section \ref{practice:awareness}, how organizations have gradually established more structured RAI practices through new processes and facilitated their employees in leveraging RAI interventions (Section \ref{practice:professionalized} and \ref{practice:RAI-interventions}), and how organizations have increasingly engaged with external stakeholders in conducting RAI work (Section \ref{practice:external}.

\subsection{Increased Awareness of RAI's Existence and Importance} \label{practice:awareness}

Across the literature, there is a clear temporal shift: more recent studies increasingly report that industry practitioners—even those not directly responsible for RAI tasks—express awareness of RAI as an important organizational priority. In line with prior work \cite{ximenes2021concrete, lee2024don, pant2024ethics, khan2023ai, das2022security}, we define the awareness of RAI as practitioners’ recognition of the existence, importance, and organizational relevance of RAI concepts and values, \textit{regardless} of whether they are directly responsible for implementing them; RAI knowledge involves the concrete skills and trainings needed to address
those issues in practice; RAI attitudes reflect the extent to which practitioners and organizations actually value RAI principles relative to competing priorities such as speed to market or model
accuracy. \looseness=-1

\subsubsection{\textbf{Overall trends of increased RAI awareness}}

Studies published between 2019 and 2022 often found that many developers lacked RAI awareness, as they had not yet recognized concepts such as AI fairness, explainability, or trustworthiness as relevant to their work prior to participating in the studies \cite{holstein2019improving, orr2020attributions, passi2019problem, lee2021landscape, rakova2021responsible, madaio2020co, widder2022limits, winecoff2022artificial, wang2022whose}. Some even reported that they believed there are no RAI-related issues at all arising from their work \cite{rakova2021responsible, wang2022whose}. Privacy was 
the main concern that was raised due to regulations such as the General Data Protection Regulation (GDPR) and California Consumer Privacy Act (CCPA), but typically in relation to software, generally, rather than AI, specifically \cite{orr2020attributions}.

In contrast, more recent studies explicitly show that participants, whether or not they are in designated RAI roles, are more aware of and able to articulate the importance of RAI (e.g., \cite{oldfield2024technical, akbar2024trustworthy, deng2023investigating, ryan2024ai, roman2024wasn, widder2024power, constantinides2024rai, habiba2025ml, lanne2025organisational, yildirim2023investigating, elsayed2023responsible, bughin2024doing, lancaster2024s}). Several survey-based papers provide quantitative evidence suggestive of this trend. For example, in 2024 Oldfield et al.  \cite{oldfield2024technical} conducted a survey comparing practitioner perspectives on fairness with those reported in 2019 by Holstein et al. \cite{holstein2019improving}. 
While acknowledging differences in sampling and study design, they found that a larger share of respondents in their study reported perceiving fairness as an important concern in AI/ML modeling work, which they interpret as indicative of “a general move within the practitioner community to prioritize fairness.” 
Similarly, surveys conducted by \citet{akbar2024trustworthy} (2024)  and \citet{bughin2024doing} (2024) both reported broad agreement among practitioners regarding the existence of and their company's policies and guidelines on addressing RAI,
as well as widespread acknowledgment of the challenges of implementing it effectively. Both author groups interpreted this as evidence of progress compared to findings in earlier studies in which practitioners were not aware of RAI concepts \cite{akbar2024trustworthy, bughin2024doing}.

In addition to quantitative results, many interview studies also document this increasing awareness through direct quotes from practitioners \cite{sadek2024challenges,vaast2025experiencing, halme2024making, schiff2024emergence, lancaster2024s, bruun2025coordination}. For instance, \citet{vaast2025experiencing} (2025) conducted a follow-up round of interviews in 2023 with the same 26 practitioners originally interviewed in 2020. This comparison revealed that practitioners perceived a broader shift toward treating AI ethics as an important issue compared to before, regardless of their different levels of experience working in RAI. Similarly, in interviews conducted by Sadek et al. (2024) \cite{sadek2024challenges}, practitioners across roles—including data scientists, AI engineers, and UX designers—emphasized the importance of creating ethical AI. Many also noted their awareness of company efforts to develop tools and guidelines for fairness and transparency, even when they themselves had not previously worked directly on RAI issues. Halme et al. (2024) \cite{halme2024making}, Schiff et al. (2024) \cite{schiff2024emergence}, Lancaster et al. (2024) \cite{lancaster2024s}, and Bruun et al. (2025) \cite{bruun2025coordination} all explicitly reported practitioners’ self-assessments of increased awareness of RAI concepts such as fairness and transparency over the years, compared to when they first began working on AI, further confirming the trend of rising awareness in recent years.

\subsubsection{\textbf{Factors shaping RAI awareness}}

Adding nuance to this overall trend, we found several factors that influence not only the degree of awareness but also how practitioners’ awareness is formed. These include regional regulations, company policies, local cultural norms, and demographic factors.

To start with, we found that \textbf{regional regulations and company policies play a significant role in shaping RAI awareness}. For many practitioners based in the U.S. and U.K.—the two largest populations represented in our corpus—\textbf{company-wide policies and public discourse} about RAI often served as the entry point for awareness \cite{pant2024ethics, kallina2025stakeholder, ryan2024ai, browne2024tech, ali2023walking, rakova2021responsible, deng2023investigating}. For example, multiple interview studies include direct quotes from practitioners who described realizing the importance of RAI when their companies began rolling out internal policies prioritizing RAI concepts \cite{pant2024ethics, kallina2025stakeholder, deng2023investigating}.
By contrast, studies with European practitioners show consistently higher awareness, with participants frequently citing exposure to or engagement with EU regulations. Professionals in Germany \cite{cociancig2024ai, sloane2022german, hedlund2025distribution, anderson2024evaluating, habiba2025ml}, the Netherlands \cite{hedlund2025distribution, bruun2025coordination, arbelaez2024integrating}, Italy \cite{voria2025fairness, arbelaez2024integrating, hollanek2025toolkit}, France \cite{anderson2024evaluating}, Denmark \cite{hedlund2025distribution, elsayed2023responsible, bruun2025coordination, inie2023summon}, Norway \cite{akbar2024trustworthy, wang2024operationalizing, papagiannidis2023toward}, and Finland \cite{akbar2024trustworthy, khan2023ai, hartikainen2023towards, akbarighatar2024responsible} often described their familiarity with RAI concepts—particularly “privacy and security”—as stemming from the GDPR \cite{khan2023ai}. Outside the EU, Pant et al. (2024) found that Australian practitioners’ awareness was shaped by the Australian Government’s AI Ethics Principles. Other studies of Australian practitioners similarly document how these principles influenced awareness of fairness, transparency, explainability, accountability, and contestability \cite{sanderson2023ai, ruster2025gaps}.

\textbf{Cultural norms} also affect practitioners’ awareness. Studying German practitioners, Sloane et al. (2022) \cite{sloane2022german} found that “cultural legacies,” such as Germany’s tradition of worker councils, helped institutionalize ethical frameworks. Participants described cultural norms that emphasized equity and privacy. Similarly, Akbar et al. (2024) report that Norwegian practitioners highlighted the importance of equity in AI outputs and incorporating user voices in design, reflecting the country’s participatory design traditions \cite{akbar2024trustworthy}. By contrast, several studies of Asian practitioners found that many technically trained professionals expressed enthusiasm for technological progress while downplaying the importance of RAI, often citing organizational cultures that prioritize innovation over ethics \cite{behl2023role, sambasivan2021everyone, thenral2021challenges, di2023ethical, vaast2025experiencing, chen2022practitioners}. Di (2023) explicitly compared Chinese and U.S. practitioners, finding that while U.S. practitioners worried about big data reinforcing racial and gender inequalities, Chinese practitioners were more optimistic, framing big data and AI as a tool to reduce income disparities across regions, showing low awareness of RAI work \cite{di2023ethical}. The broader cultural acceptance of government use of big data and surveillance technologies, including facial recognition, also appeared to contribute to relatively low RAI awareness around privacy \cite{di2023ethical}.

Although geographic, cultural, and demographic factors clearly shape RAI awareness, the participant pool in our reviewed studies remains limited in diversity. As discussed in Section \ref{paper overview}, more than half of studies recruited participants from the U.S. (38.42\%) and U.K. (14.21\%). Among the 43\% of papers that reported participants’ gender and race, many explicitly noted an overrepresentation of Western, white, cisgender, and male practitioners relative to the global distribution of AI practitioners and the populations affected by deployed AI systems. This skew aligns with recent critiques of the RAI field’s demographic skew \cite{septiandri2023weird}, which constrains the range of organizational contexts and perspectives represented in the current RAI literature.
We therefore underscore the need for broader recruitment strategies to better capture how regional and demographic factors influence practitioners’ conceptualizations of RAI. We expand on this point in our discussion of implications for future RAI research in Section \ref{dis:research}.

\subsubsection{\textbf{Different prioritization of RAI concepts}}

Four papers in our corpus explicitly compared practitioners’ prioritization of different RAI concepts \cite{jakesch2022different, khan2023ai, pant2024ethics, olson2025speaks}. Across these surveys, privacy was consistently ranked as the most important concern, aligning with the strong influence of existing privacy regulations on practitioner attention. Beyond privacy, however, findings diverged.  \citet{pant2024ethics} (2024) and \citet{olson2025speaks} (2025) reported lower practitioner prioritizations of concepts such as fairness and transparency relative to reliability, safety, accountability, and accessibility, whereas \citet{jakesch2022different} (2022) and \citet{khan2023ai} (2023) found fairness and transparency to be among the more highly prioritized concepts. Rather than representing direct contradictions, these differences likely reflect variation in study contexts and sampling, a limitation acknowledged by all four papers. 

Finally, the reviewed papers show that \emph{practitioners} often have different priorities when it comes to RAI issues than other stakeholders such as end users \cite{chen2022practitioners, jakesch2022different}. For example, \citet{jakesch2022different} compared the views of AI practitioners, a representative sample of the US population, and a sample of crowdworkers, and found that the general public overall prioritizes transparency, accountability, and privacy more than practitioners .
This result is largely in line with the qualitative evidences from interview studies conducted by Chen et al. (2022) when comparing values between practitioners and end users \cite{chen2022practitioners}. In addition, \citet{jakesch2022different} also found that AI practitioners in their study were more likely to prioritize fairness over performance, while the general public and crowdworkers were more inclined to prioritize safety and privacy over other RAI concepts.

Despite the trends of growing awareness of RAI, a large number of papers in our corpus suggested that increased RAI awareness does not necessarily imply correspondingly increased RAI knowledge and training \cite{rakova2021responsible, madaio2022assessing, widder2023dislocated, ali2023walking, lanne2025organisational}, which we expand on in Section \ref{challenge:lack of expertise}.
Similarly, although practitioners may recognize RAI values in principle, this recognition does not always translate into prioritization in day-to-day practice \cite{widder2022limits, xivuri2023ai}. Prior work suggests that such gaps between stated values and enacted priorities are often shaped by organizational context, which we expand on in Section \ref{challenge:organizational dynamics}.

\subsection{From Ad-hoc Advocacy to Professionalized Roles and Practices} \label{practice:professionalized}

In earlier studies, RAI practices were often described as \textbf{ad hoc and improvised, typically initiated by individual advocates} \cite{metcalf2019owning, holstein2019improving, passi2018trust, passi2019problem, lee2021landscape, richardson2021towards, madaio2020co, hong2020human, van2020hiring}. By contrast, more recent work (2023–present) reveals a clear shift toward the \textbf{formalization and standardization of RAI practices}. In addition, organizations have increasingly \textbf{established specialized RAI roles} aligned with concrete responsibilities, as evidenced by studies that report dedicated positions such as RAI software engineers or data scientists \cite{minkkinen2023co, rismani2023does, halme2024making}, RAI designers or UX researchers \cite{wang2023designing, deng2023understanding, schor2024mind}, AI ethics specialists \cite{reuel2025responsible, hanschke2024data, cociancig2024ai, madaio2024learning}, and RAI policy analysts \cite{wehrens2023ethics, kapania2025examining}.

In the remainder of this section, we illustrate the ongoing—though still incomplete—professionalization of RAI roles and practices through three representative RAI tasks
that are \textit{most commonly explored in the literature}: (1) AI fairness testing, (2) designing for model interpretability and transparency, and (3) documentation for RAI.
Notice that in this section, we touch on some of the tooling and processes to support these activities, but we expand on this topic in section \ref{practice:RAI-interventions}.

\subsubsection{\textbf{Case 1: AI Fairness Testing}} To start with, in early studies, many participants reported that they had not conducted any form of fairness testing in their roles \cite{metcalf2019owning, holstein2019improving, vakkuri2019ethically, hohman2019gamut, vakkuri2020current, lee2021landscape, deng2022exploring, passi2019problem}. Among those who did, fairness testing\footnote{Here, we use fairness testing to refer broadly to practices aimed at assessing whether AI systems produce systematically disparate outcomes across demographic or social groups, and to identifying and mitigating potential sources of bias in data, models, or deployment contexts \cite{barocas2023fairness}.} was typically performed by model developers in response to public incidents or driven by individual advocates, rather than being part of a formalized process \cite{metcalf2019owning, chen2022practitioners, winecoff2022artificial}. 
Several studies documented that AI practitioners often neglected fairness considerations during the initial design of AI features \cite{passi2019problem, madaio2020co}, reacted only after external reports of bias rather than proactively assessing or mitigating gender- and race-based disparities \cite{holstein2019improving, van2020hiring}, and were at times explicitly instructed by their organizations to “respond to, and ideally avoid, public calamities” instead of actively identifying harmful biases \cite{metcalf2019owning}. For those who were self-motivated to conduct fairness testing, the process was largely ad hoc and improvised \cite{deng2022exploring, richardson2021towards, madaio2022assessing}. Prior research describes instances where developers discovered potential biases “by happenstance” \cite{madaio2020co} or initiated informal fairness assessments through “water cooler conversations” \cite{deng2022exploring}.

A small number of studies have also specifically examined domains regulated by existing privacy and anti-discrimination laws and policies, such as finance, human resources, and healthcare \cite{park2022designing, thenral2021challenges, lai2020perceptions}. 
These studies consistently revealed that while organizations often attempt to \textit{signal} a commitment to fairness testing by including it as a formal requirement, they rarely provide systematic support or assign personnel with the necessary expertise to carry out such testing—resulting in practices that remain largely ad hoc despite external regulatory pressures \cite{park2022designing}.

In contrast, many practitioners in studies published after 2023 reported having more established in-house fairness testing procedures \cite{smith2025pragmatic, smith2023many, voria2024expectation, voria2025fairness, wang2024operationalizing, inie2023summon, bar2024ai}. For example, \citet{smith2025pragmatic} documented that many industry AI practitioners specializing in AI fairness had already developed dedicated pipelines for fairness testing
and had begun devising pragmatic strategies to navigate the organizational dynamics and barriers that can hinder such processes. Similarly, \citet{voria2024expectation} found that fairness testing has become “habitual” for some developers, especially when their organizations explicitly support learning and adoption of fairness toolkits. Compared to \citet{passi2019problem}’s earlier observation that fairness was often neglected during the problem formulation stage, recent studies (2024–2025) have shown a shift toward earlier and more systematic consideration of fairness. Research by \citet{wang2024operationalizing, inie2023summon, olson2025speaks, bar2024ai, browne2024tech} illustrates that many AI teams now proactively assess potential biases in training data, sensitive attributes, and anticipate disparate impacts on marginalized communities even during the initial stages of problem formulation. Practitioners from non-technical backgrounds, such as UX researchers, are also increasingly being engaged in fairness testing, bringing qualitative data from end users to surface representative harms \cite{smith2023scoping, olson2025speaks, bar2024ai}.

The professionalization of fairness testing has also coincided with the growing adoption of open-source fairness toolkits and the development of in-house fairness tools and frameworks. We further discuss the adoption of fairness toolkits in Section \ref{practice:RAI-interventions} and highlight the persisting challenges associated with these practices in Section \ref{challenge:tailored interventions}.

\subsubsection{\textbf{Case 2: Designing for model explainability and interpretability}} 

Across multiple studies published before 2022, practitioners noted that interpretability work for AI models was not yet formalized as a routine practice, but instead was often improvised by team members without clear guidance or organizational support \cite{hong2020human, hohman2019gamut, van2020hiring, benjamins2019responsible, costanza2022audits, lai2020perceptions, thenral2021challenges}. For example, based on interviews with 22 industry practitioners experienced in explainable AI, \citet{hong2020human} found that practitioners across roles frequently devised ad hoc methods to enhance model interpretability in the absence of structured processes or resources. 
Similarly, when evaluating Gamut, a prototype designed to support data scientists in addressing interpretability challenges, \citet{hohman2019gamut} reported that practitioners emphasized the need for in-house processes and tools to support model interpretation—both for generating hypotheses about data and models and for communicating insights to diverse stakeholders. 

In contrast, numerous studies published since 2023 suggest that the design of explainable AI systems is increasingly becoming a deliberate and repeatable organizational practice, rather than an ad hoc effort confined to individual teams \cite{akbar2024trustworthy, habiba2025ml, schiff2024emergence, deng2023investigating, wang2023designing, ehsan2024seamful, bruun2025coordination}. For instance, \citet{akbarighatar2024operationalizing} found that organizations have begun establishing explicit goals on transparency and explainability for their AI models. Similarly, \citet{habiba2025ml} reported that German industry practitioners described formalized requirements and practices for model explainability, shaped by regulatory obligations, client and business demands, corporate risk management, and internal ethical policies. This formalization is also reflected in the growing use of internal audits to assess model transparency, often driven by emerging regulatory frameworks at international, national, and local levels \cite{schiff2024emergence}. Importantly, recent studies in high-stakes domains such as healthcare and finance also suggest that practitioners are now more often explicitly tasked with designing explainability features and collaborating with domain experts to ensure their effective use \cite{wehrens2023ethics, nichol2023developer, nichol2024moral}.

Finally, while most of this research focuses on technical roles, recent studies have also illuminated the critical contributions of non-technical roles in explainability and transparency work. Through a series of interview and workshop studies, \citet{deng2023investigating}, \citet{wang2023designing}, \citet{ehsan2024seamful}, and \citet{bruun2025coordination} showed how UX professionals and program managers often act as coordinators, leveraging cross-disciplinary expertise to facilitate collaboration around RAI tasks such as model transparency. For instance, \citet{wang2023designing} described how UX practitioners translated user insights from interactions with large language models (LLMs) into actionable feedback for engineers—helping design more interpretable model outputs and providing users with greater agency in steering responses. However, significant challenges remain in standardizing and improving transparency design practices. Studies have repeatedly shown that practitioners may misinterpret or mistrust interpretability analyses \cite{kaur2020interpreting, hong2020human, ashtari2023discovery, kaur2024interpretability}. We elaborate on the ongoing challenges in Section \ref{challenge}.

\subsubsection{\textbf{Case 3: RAI documentation}}

Between 2018 and 2022, many studies reported that documentation processes for RAI activities were frequently absent, largely due to limited organizational awareness of their importance and incentives to actually document RAI activities \cite{madaio2020co, andrus2021we, passi2019problem, holstein2019improving, heger2022understanding, chang2022understanding}.  
In line with prior work \cite{chang2022understanding, heger2022understanding}, we use RAI documentation to refer to the process of recording the design, development, deployment, and evaluation details relevant to RAI principles and values. The absence of such documentation was often found to hinder knowledge transfer across teams and to create frustration among industry practitioners \cite{chang2022understanding, passi2018trust, heger2022understanding, dhanorkar2021needs}. For instance, through a field study of industry practitioners, \citet{passi2018trust} found that insufficient documentation of feature engineering and metric selection led to ambiguities in evaluation results and difficulties in explaining model outputs. Similarly, \citet{heger2022understanding} found that data documentation efforts among practitioners were often piecemeal and narrowly scoped, with little connection to RAI goals such as fairness or transparency.

More recent studies, however, suggest that documentation for RAI purposes has become a more standardized and institutionalized activity, often carried out by designated professionals. 
These activities range from defining high-level RAI goals \cite{ryan2024ai, xiao2025might}, to improving diversity in data annotation \cite{gutierrez2025explaining, bhattacharya2025explanatory}, to evaluating model outputs for potential harms \cite{madaio2024tinker, smith2025pragmatic, oldfield2024technical}, and to monitoring system behavior in the wild to assess societal impacts \cite{deng2023understanding, hadley2024investigating, yildirim2023investigating}. In particular, a growing body of work documents practitioners’ efforts to implement structured documentation practices for model outputs, aimed at supporting downstream RAI activities such as fairness testing and explainability design \cite{madaio2024tinker, oldfield2024technical, smith2025pragmatic}. In addition, recent research by \citet{halme2024making, deng2023understanding, yildirim2023investigating, hadley2024investigating} reports that practitioners increasingly document user-reported incidents and model failures in real-world deployments, building internal databases to help teams understand AI system behavior across diverse contexts. Notably, several recent studies have found that producing high-quality documentation for continuous model evaluation has even become part of some practitioners’ key performance indicators (KPIs), motivating sustained engagement in documentation work despite its labor-intensive nature \cite{bhattacharya2025explanatory, madaio2024learning}. This marks a significant shift from earlier years, when documentation was typically self-motivated, informal, and rarely incentivized \cite{heger2022understanding, chang2022understanding, madaio2020co}. \looseness=-1

However, despite these advances, persistent challenges remain in scaling and sustaining RAI documentation practices. Studies highlight ongoing issues such as inconsistent documentation standards across teams \cite{deng2023investigating, balayn2025unpacking}, limited integration between documentation and other governance tools \cite{constantinides2024rai}, and unclear ownership of documentation responsibilities \cite{widder2024power}. Practitioners also report difficulties in maintaining documentation amid rapid model iteration cycles \cite{kallina2025stakeholder, xiao2025might}, and concerns that documentation may devolve into a compliance-oriented exercise rather than a meaningful process for reflection and accountability \cite{bar2024ai, lanne2025organisational}. We elaborate on these remaining challenges in Section \ref{challenge}. \looseness=-1

\subsection{Gradually Adopted RAI Policies, Processes, and Tools} \label{practice:RAI-interventions}

In line with the professionalization of RAI roles, many studies report that industry practitioners \textbf{have already begun leveraging various RAI interventions to support and improve their practices}. Following prior work \cite{berman2024scoping}, we use the term RAI interventions as an umbrella concept encompassing RAI policies, processes, and tools, where we define RAI policies as internal organizational rules and external regulations governing RAI work; RAI processes as guidelines and documentation that structure internal RAI practices; and RAI tools as computational systems that enable technical roles to carry out RAI activities.

At a high level, it is evident that \textbf{adoption of these interventions has steadily increased over the years}. In earlier studies, participants were often invited to evaluate or provide feedback on RAI tools introduced by researchers during the studies (e.g., \cite{kaur2020interpreting, lee2021landscape, deng2022exploring, thenral2021challenges, hopkins2021machine, hong2020human, madaio2020co, balayn2023fairness}). Many of them had never previously used—or even heard of—such tools or frameworks \cite{deng2022exploring, lee2021landscape, hong2020human, ehsan2021expanding}. By contrast, in more recent studies, many participants reported that they were already using these tools in practice \cite{bogucka2024co, habiba2025ml, xiao2025might, voria2025fairness, madaio2024learning, madaio2024tinker}, and adapted publicly available RAI interventions to fit their own workflow and environment \cite{schiff2024emergence, madaio2024learning, madaio2024tinker, habiba2025ml, ruster2025gaps, smith2025pragmatic, ryan2024ai, schor2024mind}. 

In the following subsections, we discuss how some commonly seen policies, processes (such as documentation and guidelines), and computational tools are increasingly being adopted and integrated into on-the-ground RAI practices within industry settings.

\subsubsection{\textbf{Internal RAI Policies}} 
Many papers suggested that companies are increasingly adopting formal internal policies to navigate evolving regulatory landscapes and institutionalize ethical standards \cite{akbar2024trustworthy, bughin2024doing, ryan2024ai, bach2025insights, bar2024ai, voria2025fairness, madaio2024learning, madaio2024tinker, smith2025pragmatic, hollanek2025toolkit}. 
Earlier literature often noted a "decoupling" between high-level corporate AI principles and the daily realities of technical teams \cite{orr2020attributions, holstein2019improving}, and in a number of studies, practitioners suggested a lack of cohesive company RAI policies \cite{metcalf2019owning, vakkuri2020just, rakova2021responsible}. However, recent studies indicate a shift toward more standardized internal company RAI guidelines and more robust governance structures \cite{akbar2024trustworthy, bughin2024doing}. For instance, \citet{akbar2024trustworthy} (2024) found that organizations are now formalizing "Trustworthy AI" frameworks that move beyond voluntary guidelines into mandatory compliance checklists and internal audits. Similarly, many works in 2024 and 2025 highlight that leading firms are integrating RAI criteria directly into their procurement and product development lifecycles, ensuring their AI tools meet specific risk thresholds before deployment \cite{ryan2024ai, bach2025insights, bar2024ai, voria2025fairness}. This institutionalization often involves the creation of dedicated RAI steering committees or cross-functional oversight boards, which serve to bridge the gap between abstract ethical goals and operational requirements \cite{madaio2024learning, madaio2024tinker, smith2025pragmatic, hollanek2025toolkit}. By embedding these policies into the organizational fabric, practitioners are finding more institutional support and clearer mandates for prioritizing fairness and transparency in their technical workflows.

\subsubsection{\textbf{RAI Processes}} 
Many processes such as guidelines and documentation have also been adopted by practitioners to guide their RAI activities. For example, \citet{yildirim2023investigating} (2023) found that the People + AI Guidebook---a practical guideline developed by Google to help designers and product teams create human-centered AI applications---has been credited by practitioners for surfacing previously overlooked risks to customers in their day-to-day work practices. In their study on the articulation work required to contextualize AI fairness checklists, \citet{madaio2024tinker} (2024) observed that practitioners in roles such as data scientists and product managers actively customized checklist items and integrated them into existing workflows and tools. This represents a notable shift from earlier findings, where fairness checklists were rarely incorporated into practitioners’ daily practices \cite{madaio2020co}. A number of papers also suggested a greater adoption of documentation such as model cards \cite{mitchell2019model}, datasheets for datasets \cite{gebru2021datasheets}, and system cards \cite{SystemCards} in their AI development \cite{schor2024mind, ali2023walking, ashtari2023discovery, bennett2025everybody, ruster2025gaps}. 

In addition to industry-driven tools and processes, various guidelines have been developed or recommended by government agencies and standards organizations. For instance, the National Institute of Standards and Technology (NIST) has released the AI Risk Management Framework (AI RMF) \cite{AIRiskManageF}, which serves as a broad guideline to help organizations ensure responsible and trustworthy AI development and deployment. Recent work by \citet{ryan2024ai} and \citet{schor2024mind} has shown that practitioners are increasingly adapting and embedding such high-level frameworks, including the NIST AI RMF, into their organizational practices to align internal processes with emerging regulatory and ethical standards.

\subsubsection{\textbf{Computational Toolkits}}
Recent literature highlights a notable shift in how practitioners engage with open-source computational toolkits designed for fairness evaluation and interpretability. Prior to 2022, studies consistently found that practitioners had limited familiarity with such toolkits \cite{lee2021landscape, deng2022exploring, thenral2021challenges, hopkins2021machine, lee2021risk}. These resources—such as Microsoft's Fairlearn \cite{FairlearnAPI, bird2020fairlearn}, IBM’s AIF360 \cite{AIF360API}, and Google’s What-If tool \cite{What-If}—were originally developed to provide technical support for Responsible AI (RAI) activities, yet they initially struggled to gain traction in industry workflows.

In contrast, research since 2023 indicates that practitioners are now actively adopting and adapting these toolkits into their daily operations \cite{deng2023understanding, schiff2024emergence, voria2025fairness, ruster2025gaps, smith2025pragmatic}. For example, \citet{deng2023understanding} and \citet{schiff2024emergence} observe that these tools are frequently utilized to audit AI products and services. This trend extends to interpretability toolkits like IBM's AIX360 \cite{AIX360API}, which has seen increased organizational use in recent years \cite{xivuri2023ai, gutierrez2025explaining}. Furthermore, \citet{habiba2025ml} (2025), among others \cite{halme2024making, kallina2025stakeholder}, found that practitioners are no longer just using model-agnostic techniques like SHAP \cite{lundberg2017unified} and LIME \cite{ribeiro2016should} out-of-the-box; instead, they are increasingly customizing these explanations to highlight the specific features and outputs most relevant to their stakeholders.

Overall, these computational tools have been widely praised by practitioners for lowering the barrier to engaging in RAI work compared to earlier years, when such efforts required implementing everything from scratch without centralized, research-driven resources \cite{schiff2024emergence, smith2025pragmatic, madaio2024learning}. However, there remain gaps in supporting industry practitioners in effectively and appropriately using these toolkits, which we discuss in Section \ref{challenge:tailored interventions}. \looseness=-1

\subsection{Attempts at External Stakeholder Engagement for RAI Work} \label{practice:external}

Responsible AI work does not end at the boundaries of a development team. Effective RAI practice requires input, feedback, and oversight across the entire system lifecycle---from early design through deployment and beyond---and often depends on perspectives from people who are not directly involved in building the system. Recognizing this, some industry practitioners have moved beyond internal processes to actively engage external stakeholders as part of their RAI efforts. 
We highlight the three most commonly seen external stakeholder engagements documented in the prior research: end users, external domain experts, and third party data annotators.

\subsubsection{\textbf{Attempts to engage end users in evaluating AI products and services}}

Prior to 2022, research indicated that while practitioners recognized the importance of engaging with AI end users---the people directly using an AI product or service---there was little evidence of this being implemented in practice \cite{metcalf2019owning, holstein2019improving, vakkuri2020just}. Since 2023, however, a growing body of scholarship has documented practitioners actively involving users across various RAI activities. For example, \citet{deng2023understanding} and \citet{wang2023designing} found that practitioners now engage end users in testing, auditing, and red-teaming to uncover edge cases that internal teams might otherwise overlook.

Beyond direct participation, recent studies highlight the development of specialized methods and infrastructures to facilitate this engagement \cite{vereschak2024trust, yuan2023contextualizing, kallina2025stakeholder, halme2024making, deng2023understanding}. For instance, \citet{halme2024making} described the use of “\textit{Ethical User Stories}”—firsthand accounts of real-world AI interactions—to inform responsible decision-making and evaluate downstream impacts. Similarly, practitioners are increasingly collaborating with customer service and trust and safety teams to integrate user-reported incidents into formal auditing processes \cite{vereschak2024trust, yuan2023contextualizing, kallina2025stakeholder}.

Collectively, these findings reveal that end-user involvement remains largely indirect---often mediated by user-facing roles like UX researchers---and typically occurs in the later stages of development after core features are finalized \cite{yildirim2023investigating, wang2023designing, sadek2024challenges}. Consequently, end-user engagement is primarily relegated to post-hoc evaluation and issue detection rather than shaping the foundational design or governance of AI systems. 
We further detail the challenges of involving users earlier and throughout the entire AI lifecycle in Section \ref{challenge:external friction}.

\subsubsection{\textbf{Engaging external domain experts}}

An increasing number of studies have documented practitioners’ efforts to collaborate with external domain experts to access specialized knowledge needed to address domain-specific RAI concerns that would otherwise be difficult to identify and resolve within their organizations. \cite{oldfield2024technical, pink2024trust, morley2023operationalising, arbelaez2024integrating, nichol2023developer, labkoff2024toward, nichol2024moral, bhattacharya2025explanatory}. For instance, research on AI development in healthcare has shown that practitioners actively engage clinicians, medical ethicists, and other healthcare professionals to evaluate AI systems and ensure clinical relevance and safety \cite{oldfield2024technical, pink2024trust, morley2023operationalising, arbelaez2024integrating}. \citet{arbelaez2024integrating}, for example, found that collaborations between industry practitioners and healthcare experts often reveal tensions between commercial and medical priorities, with experts expressing concern that product-driven norms---such as speed to market and profitability---conflict with medical ethics and patient care standards. A number of works also documented similar dynamics in other regulated domains, such as finance \cite{passi2019problem, sadek2024challenges}, retail \cite{lee2024don, xivuri2023ai, morley2023operationalising}, and insurance \cite{cociancig2024ai, nichol2023developer, habiba2025ml}, when external domain experts were brought into the RAI work. 

In addition, in light of the existing regulation such as GDPR and CPAA, as well as the forthcoming AI regulations such as EU AI Act, recent years have seen increasing involvement of policy and regulatory experts---often from civil society, academia, and government agencies---in industry RAI discussions \cite{hollanek2025toolkit, bogucka2024co, halme2024making}. For example, \citet{hollanek2025toolkit} found that some practitioners have begun collaborating with policy experts, including those involved in drafting the EU AI Act, to ensure that their responsible AI toolkits align with emerging regulatory requirements and compliance frameworks.

Despite these promising developments, collaborations with external stakeholders---including annotators, end users, and domain experts---continue to face significant frictions. Challenges often arise around power asymmetries, lack of communication infrastructure, and limited incentives for sustained collaboration. We further discuss these barriers---and potential strategies to address them---in Section \ref{challenge:external friction}.

\subsubsection{\textbf{Progress on managing third party data annotators}}
Data annotators play critical (yet often invisible) role in AI development as they are the one providing ground truth for dataset used for model training and the evaluation \cite{miceli2020between}.
A small but growing body of research has examined how industry AI teams manage and collaborate with third-party data annotators \cite{kazimzade2020biased, wang2022whose, sambasivan2021everyone, sambasivan2022deskilling, kapania2023hunt}.
Early studies painted a rather negative picture: practitioners often regarded data annotation as a less prestigious task, delegating it to third-party vendors with limited oversight or appreciation of annotators’ expertise \cite{sambasivan2021everyone, sambasivan2022deskilling}. As a result, annotators’ skills, contextual knowledge, and even their emotional well-being were frequently overlooked. These studies also highlighted how annotators---especially those in the Global South---experienced deskilling, poor working conditions, and limited recognition despite their critical role in shaping model behavior.

More recent research, however, suggests that this landscape is beginning to shift. Through a combination of surveys and interviews, \citet{kapania2023hunt} found increasing awareness among AI practitioners of the importance of ensuring diversity among data annotators to improve dataset representativeness and fairness. Some practitioners have also begun developing internal guidelines to ensure equitable compensation and better working conditions for annotators \cite{kazimzade2020biased, wang2022whose}. For example, several industry teams have experimented with revising pay structures, providing mental health resources, and improving communication channels between engineers and annotation teams to foster mutual understanding and accountability \cite{sambasivan2021everyone, sambasivan2022deskilling}.

\section{Persistent Challenges and Opportunities in Doing RAI in Industry} \label{challenge}

The previous section presented current practices and trends in industry RAI, shedding light on improvements in RAI practices surfaced through our literature review. However, despite these reported “success stories,” persistent challenges continue to be articulated by industry practitioners across the literature. In this section, we synthesize these challenges and, within each subsection, present corresponding opportunities for addressing them as discussed in the papers. These opportunities are often directly articulated by practitioners, but may also reflect suggestions or implications proposed by researchers. We also highlight points of disagreement among authors where differing perspectives on opportunities emerge. Figure~\ref{fig:result overview} provides an overview of the identified challenges and opportunities, mapped to the practices we observed.

\subsection{Lack of Sufficient RAI Knowledge and Training} \label{challenge:lack of expertise}

Despite the increased awareness of RAI’s existence and importance among industry practitioners (see Section \ref{practice:awareness}), as well as the increasing adoption of RAI tools (see section \ref{practice:RAI-interventions}), a large body of work suggests that this awareness does not necessarily translate into deep or actionable knowledge. In particular, more than 120 papers over the years find that even practitioners tasked with RAI work often lack sufficient expertise and training in relevant technical and socio-technical domains to carry it out effectively. 

To start with, practitioners often lack \textbf{sufficient technical knowledge to effectively use these RAI tools in their work} \cite{sarathy2023don, lee2024don, lee2021landscape, richardson2021towards, deng2022exploring, cinca2025practitioners, hollanek2025toolkit, madaio2024learning, voria2025fairness}. This phenomenon is particularly pronounced in fairness testing. In particular, multiple studies surfaced that practitioners often \textbf{lack knowledge on the mathematical underpinning} behind commonly used fairness metrics and bias mitigation techniques, which impede their effective use of these toolkits  \cite{lee2021landscape, richardson2021towards, deng2022exploring, cinca2025practitioners}. Throughout years, interviews with industry practitioners, ranging from \citet{lee2021landscape} (2021), to \citet{richardson2021towards} (2021), \citet{deng2022exploring} (2022) and \citet{cinca2025practitioners} (2025), all surfaced that practitioners who desire to use fairness toolkits often do not understand crucial differences between bias metrics, such as disparate impact and equal opportunity, or the differences between bias mitigation strategies, such as in-process and post-process. Because of this lack of fundamental knowledge on AI fairness, practitioners across many interview studies reported a \textbf{steep learning curve} in using fairness tools, often due to these toolkits (at the time) not providing a sufficient explanation of these metrics and techniques \cite{lee2021landscape, hollanek2025toolkit, madaio2024learning}. 

Similarly, in the area of privacy testing, \citet{sarathy2023don} (2023) found that practitioners often lack a basic understanding of techniques such as differential privacy and federated learning, which prevents them from applying appropriate methods to ensure privacy in practice. This challenge is further exacerbated by the lack of intersectional expertise in privacy and AI among practitioners, as highlighted by \citet{lee2024don} (2024). A number of surveys provide evidence that a lack of knowledge and training on RAI topics is often the top challenge for practitioners to carry out RAI work \cite{pant2024ethics, olson2025speaks, akbar2024trustworthy, voria2025fairness, lu2024responsible, xivuri2023ai}.

Due to insufficient knowledge around RAI mentioned above---and the limited resources available to address these knowledge gaps---practitioners sometimes \textbf{even misuse RAI tools or misinterpret their results} \cite{kaur2024interpretability, ashtari2023discovery, habiba2025ml, bhattacharya2025explanatory, deng2022exploring, balayn2023fairness, hanschke2024data}. For instance, \citet{kaur2020interpreting} (2020) observed that data scientists misused interpretability tool outputs by over-trusting visualizations they did not fully understand, leading them to wrongly rationalize suspicious model behavior. This pattern of misinterpretation and over-trust has repeatedly surfaced in more recent studies published between 2023 and 2025 \cite{kaur2024interpretability, ashtari2023discovery, habiba2025ml, bhattacharya2025explanatory}. In the context of fairness testing, \citet{deng2022exploring} (2022) found that participants who assumed sex was a sensitive feature attempted to mitigate bias by simply removing or ignoring such features during preprocessing, an approach that neglects well-established “fairness through awareness” principles in AI fairness research \cite{barocas2023fairness}. Similarly, \citet{balayn2023fairness} (2023) reported that many practitioners applied all fairness metrics available in toolkits without understanding their meanings or appropriateness, attempting to falsely justify a model’s fairness using the metrics they picked. A growing body of recent work has echoed these findings, showing that limited technical understanding of AI fairness and interpretability remains one of the most common challenges practitioners encounter when conducting RAI work that led to potential misuses of RAI tools \cite{hanschke2024data, voria2025fairness, lu2024responsible, xivuri2023ai}. 

Finally, \textbf{the lack of RAI knowledge can hinder effective communication and collaboration across roles}. Numerous studies report that \textit{technical roles}---such as data scientists, software engineers, and machine learning engineers---often lack training in conveying the results of fairness audits or presenting interpretability results from SHAP or LIME to \textit{non-technical roles}, including designers, business managers, and compliance staff \cite{hine2024ethics, madaio2024tinker, kommiya2024towards}. Conversely, practitioners with backgrounds in ethics, law, and design often struggle to translate RAI policies or design principles into actionable guidance for technical teams \cite{dhanorkar2021needs, sanderson2023ai, schor2024mind, smith2023scoping, deng2023investigating, madaio2024tinker}. Interview studies throughout years conducted by \citet{dhanorkar2021needs} (2021), \citet{park2022designing} (2022), \citet{sanderson2023ai} (2023), \citet{deng2023investigating} (2023) and \citet{schor2024mind} (2024) consistently show that many product managers and compliance professionals felt unprepared to communicate high-level AI governance goals in ways that could be concretely implemented within AI development pipelines. Many reported that the skills they developed in management or policy work did not directly transfer to addressing responsible AI issues in team settings. In several workshop studies, practitioners such as UX researchers and product managers also expressed frustration when attempting to establish design requirements or roadmaps for AI fairness testing, noting that they often underestimated the required resources and technical limitations involved \cite{smith2023scoping, deng2023investigating}. 

Interestingly, we see growing evidences that \textbf{the rise of generative AI has further exacerbated these knowledge and training gaps} \cite{wang2023designing, liao2023designerly, hadley2024investigating, xiao2025might, smith2025pragmatic}. Recent studies show that practitioners struggle to define and assess “fair output,” particularly given the vast output space and ambiguously impacted groups associated with generative models \cite{wang2023designing, liao2023designerly, hadley2024investigating}. For example, \citet{wang2023designing} (2023), \citet{liao2023designerly} (2023), and \citet{hadley2024investigating} (2024) all found that non-technical roles such as UX researchers and policy officers reported that generative AI systems feel more opaque than prior classification-based models, requiring additional technical understanding to evaluate and govern effectively. Roles such as UX researchers and designers often struggle to design for model transparency due to limited understanding of model capabilities themselves, particularly when working with pre-trained general-purpose model \cite{liao2023designerly, ehsan2021expanding}. 

\subsubsection{\textbf{Opportunities for Supporting RAI Learning and Training before and on the Job}}  \label{opp:learning}

Given that these knowledge gaps hinder multiple aspects of RAI work, many papers propose education and training as key levers for addressing these deficits \cite{rismani2023does, deng2023investigating, browne2024tech,lu2022software, akbar2024trustworthy, pant2024ethics}.

To start, a number of studies call for \textbf{incorporating RAI education and training into current AI-related curriculum} to keep pace with the rapidly evolving AI and RAI landscape \cite{rismani2023does, deng2023investigating, browne2024tech, ryan2023integrating, agbese2023implementing, voria2024expectation, bennett2025everybody, de2024preliminary, gutierrez2025explaining, wang2024operationalizing, lu2022software, akbar2024trustworthy, pant2024ethics, dominique2023factsheets}. For example, drawing on interviews with RAI industry experts, \citet{rismani2023does} (2023) synthesized an ontology of RAI roles—spanning technical research, data science, engineering, policy, design, social science, and management—to identify the knowledge and skills each role requires. They argue for curricula that better encompass these knowledge bases when training future AI practitioners \cite{rismani2023does}. To address collaboration challenges rooted in misunderstandings of RAI as purely technical work, many papers also call for AI and computer science education to integrate AI ethics into the curriculum, sensitizing practitioners to the socio-technical and human-centered nature of responsible AI before they enter the workforce \cite{deng2023investigating, browne2024tech}. To this end, a number of papers called for universities to prepare students for roles in which they could serve as a liaison between technical and non-technical stakeholders to ensure shared understanding as well as effective implementation of RAI \cite{ryan2023integrating, agbese2023implementing, voria2024expectation, bennett2025everybody, de2024preliminary, gutierrez2025explaining, wang2024operationalizing, lu2022software, akbar2024trustworthy, pant2024ethics, dominique2023factsheets}. Interestingly, work in software engineering venues has highlighted opportunities to expand “requirements engineering," a must-learn technique in many software design classes, to explicitly cover core responsible AI concepts such as fairness and transparency \cite{lu2022software, akbar2024trustworthy, pant2024ethics, dominique2023factsheets}.

Research also acknowledges that not all the knowledge and training needed for real-world RAI work can be acquired in formal education, and advocate \textbf{creating opportunities to support RAI learning on the job}  \cite{madaio2024learning, deng2023investigating, voria2024expectation, bennett2025everybody, de2024preliminary, hanschke2024data, bogucka2024co, boyd2021datasheets, roman2024wasn, drage2024engineers, bar2024ai, smith2023many, smith2025pragmatic, elsayed2023responsible, ryan2023integrating, agbese2023implementing}.  In particular, \citet{madaio2024learning} (2024) document the growing phenomenon of “learning about RAI on the job,” which enables practitioners to develop skills through mentorship, peer support, and communities of practice. They call for companies to invest in mentoring structures and internal communities to facilitate this interpersonal learning \cite{madaio2024learning}. In addition, a large body of work also highlights the importance of companies providing internal training and education programs to support RAI practitioners’ learning on the job \cite{deng2023investigating, voria2024expectation, bennett2025everybody, de2024preliminary, hanschke2024data, bogucka2024co, boyd2021datasheets, roman2024wasn, drage2024engineers, bar2024ai, smith2023many, smith2025pragmatic, elsayed2023responsible, ryan2023integrating, agbese2023implementing}. These papers call for organizations to intentionally train practitioners to work effectively in cross-functional, rapidly evolving environments \cite{smith2023many, deng2023investigating, voria2024expectation, bennett2025everybody, de2024preliminary, hanschke2024data, bogucka2024co}, and to invest in organizational learning efforts that help teams across roles stay informed about the shifting landscape of RAI regulations \cite{bogucka2024co, roman2024wasn, bar2024ai, smith2025pragmatic, elsayed2023responsible, ryan2023integrating, agbese2023implementing}.

Finally, recognizing that increasing practitioners’ knowledge or training alone cannot fully close the gaps in RAI practice, recent work has called for \textbf{developing interventions that directly support cross-role collaboration and help bridge knowledge gaps}. For instance, we see encouraging role-targeted communication tools: \citet{elsayed2023responsible} (2023) and \citet{constantinides2024rai} (2024) introduced guidelines and visualization strategies to translate technical evaluation outcomes into formats accessible to non-technical decision-makers. Other work has developed tailored dashboards with visualizations accessible to non-technical practitioners \cite{hohman2019gamut, wang2024farsight} (2019), narrative explanations of RAI workflow \cite{widder2024power}, and modular reporting templates or checklists \cite{hollanek2025toolkit, bhattacharya2025explanatory, madaio2020co} to help technical teams adapt their outputs for diverse audiences, addressing persistent communication gaps between technical and non-technical roles.

\subsection{Organizational Dynamics and Misalignment} \label{challenge:organizational dynamics}

Perhaps the most salient challenge of industry RAI, mentioned by nearly all papers in the corpus, is the general lack of organizational support for RAI efforts compared to efforts to innovate and deploy AI systems. In particular, in our corpus of 161 papers, \textbf{more than 120 papers} throughout the years have mentioned that RAI work often is not prioritized by the company compared to other AI development work. From early work done by \citet{holstein2019improving} (2019), \citet{orr2020attributions} (2020), and \citet{rakova2021responsible} (2021) to more recent work done by \citet{ryan2024ai} (2024) and \citet{bennett2025everybody} (2025), the literature consistently reports that the profit-driven and fast-paced organizational dynamic within industry has often led to resource constraints for doing RAI work. 

Despite the gradual professionalization of RAI work (Section~\ref{practice:professionalized}), practitioners consistently report significant \textbf{human resource constraints}. Across many studies, RAI teams are described as severely understaffed \cite{ashtari2023discovery}. In particular, the literature repeatedly highlights a lack of design-focused, user-facing, and policy-oriented roles, despite their recognized importance for addressing RAI challenges \cite{ghatar2023practices, papagiannidis2023toward, nakao2023towards, deng2023investigating, olson2025speaks}, and when such roles do exist, they are often fragile and lack job security \cite{widder2024power, rakova2021responsible, ali2023walking}. On the technical side, studies note that software engineers and data scientists are frequently prioritized for core model development, leaving limited technical capacity for fairness testing, interpretability, and other RAI activities that also require specialized expertise \cite{deng2023investigating, ashtari2023discovery, ruster2025gaps, xiao2025might}. 
RAI practitioners who are mid- and senior-level leaders within organization often allocate hiring and staffing resources toward product and model development rather than RAI, reinforcing these constraints \cite{ibanez2022operationalising, rismani2023does}. Although more recent work (2024–2025) indicates some improvement through the emergence of dedicated RAI roles \cite{roman2024wasn, vaast2025experiencing}, practitioners still report substantial disparities in staffing levels between RAI and core AI development. These persistent human resource constraints are widely cited as a major driver of burnout in RAI work, as practitioners are often required to extend beyond their formal roles to sustain responsible AI efforts, increasing workload and stress \cite{deng2023investigating, nichol2024moral, widder2024power, madaio2024learning, madaio2022assessing, olson2025speaks}. \looseness=-1

In addition to constraints on human resources, \textbf{limited time and finance allocation} is also a major bottleneck for much RAI work. Practitioners in many studies reported that they often must operate within the timelines and budgets set by model developers or product feature teams, leaving them with unrealistically short periods to conduct meaningful evaluations of fairness or privacy-related issues \cite{smith2025pragmatic, lancaster2024s, crockett2021building, sloane2022german}. For example, participants from \citet{lancaster2024s} (2024) and \citet{smith2025pragmatic} (2025) mentioned that unreasonable time constraints were one of the key factors hindering their RAI efforts. These constraints often resulted in only surface-level evaluations of AI systems for bias, preventing them from conducting the deeper analyses they could have performed with more time \cite{smith2025pragmatic, lancaster2024s}. Many papers also suggested that when the investment in RAI will cost company money, or delay the launch of AI systems, individuals frequently encounter pushback from leadership when advocating for RAI work  \cite{widder2023dislocated, kallina2025stakeholder, smith2025pragmatic, baker2021management, winecoff2022artificial, bessen2022cost}. 

These resource constraints in RAI work often \textbf{demotivate individual practitioners} from engaging in meaningful RAI efforts, as practitioners across multiple studies described frustration---both personally and among their colleagues---stemming from issues like insufficient time, staffing, and organizational support mentioned above \cite{deng2023investigating, madaio2020co, madaio2024tinker, rakova2021responsible, orr2020attributions, ryan2024ai, sambasivan2021everyone, widder2022limits, passi2019problem, widder2023dislocated, kallina2025stakeholder, smith2025pragmatic, baker2021management, winecoff2022artificial, bessen2022cost, lanne2025organisational, vakkuri2020just, boyd2021datasheets}. Practitioners in some studies even shared that many of their colleagues choose not to work on RAI due to these issues and select work that provides them more resources \cite{rakova2021responsible, orr2020attributions, ryan2024ai, sambasivan2021everyone, widder2022limits, passi2019problem, widder2023dislocated, kallina2025stakeholder}. Even for those who continue working on RAI, many of them self-reflected on how resource constraints have prevented them from doing RAI work to their desired standard \cite{madaio2020co, vaast2025experiencing, lanne2025organisational, vakkuri2022software, boyd2022designing, richardson2021towards, madaio2022assessing, deng2023investigating, balayn2023fairness, ashktorab2023fairness, rakova2021responsible}. For example, many practitioners mentioned that they are afraid that their current processes of doing RAI in their organization risk becoming bureaucratic “check-the-box” exercises rather than reflective, substantive practices due to the resource constraints \cite{madaio2020co, vaast2025experiencing, lanne2025organisational}. Documentation is often reduced to a compliance task \cite{madaio2020co, dominique2023factsheets, chang2022understanding, winecoff2022artificial}, while fairness and explainability evaluations can become performative practices—or even be criticized as ethics washing 
\cite{richardson2021towards, madaio2022assessing, deng2023investigating, balayn2023fairness, ashktorab2023fairness}. 
These time and budget constraints also 
negatively affected work such as engaging external stakeholders, which we expand on in Section \ref{challenge:external friction}.

Interestingly, compared to large technology companies, \textbf{startups face greater challenges in securing organizational support for RAI efforts} \cite{hopkins2021machine, winecoff2022artificial, bessen2022cost, crockett2021building, vakkuri2020just, sloane2022german}. For example, \citet{crockett2021building} (2021) found that practitioners in small businesses and startups recognized the importance of conducting bias assessments or explainability work for their models. However, the survival pressures and fast-paced nature of AI development in these startup environments often leave little room for RAI activities, especially when compared to larger technology companies that have dedicated RAI teams \cite{crockett2021building, vakkuri2020just, sloane2022german}.

\subsubsection{\textbf{Opportunities to Motivate RAI Work in Industry Settings}} \label{opp:motivation}

In line with prior work investigating motivations for industry organizations doing privacy, security, and environmental sustainability work, industry practitioners shared, across a number of studies, four primary potential motivators for convincing their teams or leaderships to provide more support for RAI work: 1. Reducing the effort necessary to engage in  effective RAI work \cite{madaio2020co, deng2022exploring, wang2023designing, wang2024operationalizing, smith2023many, dominique2023factsheets, lee2021landscape, bessen2022cost, ehsan2024seamful, balayn2023fairness, voria2024expectation}; 2. focusing on long-term profit and corporate reputation \cite{chang2022understanding, lanne2025organisational, mokander2023operationalising, sloane2022german}; 3. attending to the demand on RAI from end users \cite{}; 4. complying with existing and potential future regulations \cite{minkkinen2023co, orr2020attributions, ali2023walking, schiff2024emergence, ryan2023integrating, park2022designing}. 

To directly address the resource constraints that many organizations face, a substantial body of work calls for \textbf{tools and processes that reduce the time, effort, and expertise required to conduct RAI work}. Closely related to \ref{opp:learning}, one recurring recommendation is to embed \textbf{learning resources and scaffolding mechanisms} into RAI tooling so that practitioners can more readily interpret outputs, select appropriate evaluation approaches, and act on results in real-world settings \cite{deng2022exploring, balayn2023fairness, ehsan2021expanding, kaur2020interpreting, kaur2024interpretability, lee2021risk, chazette2022can}. For instance, prior studies argue that tools should not only surface fairness or transparency metrics, but also provide \textbf{actionable contextual guidance}---such as explanations of what different metrics capture, assumptions and failure modes, suggestions of when a method is appropriate, and workflow-integrated checklists or decision aids that help teams translate diagnostic signals into concrete follow-up actions \cite{balayn2023fairness, ehsan2024seamful}. In addition, multiple studies emphasize that RAI tools and processes should be \textbf{aligned with practitioners' existing workflows}---e.g., fitting into routine model development, evaluation, and deployment processes---so that they can be adopted with minimal overhead and meaningfully improve efficiency \cite{deng2022exploring, lee2024don, xivuri2023ai}. This workflow alignment is frequently missing from academically developed RAI interventions, which are often evaluated in isolation rather than under the constraints of real-world development practices \cite{deng2022exploring, lee2024don, xivuri2023ai}. 

To address companies' broader profit goals and the financial constraints they impose on RAI work, a number of papers suggest \textbf{highlighting the long-term benefits of responsible AI practice} \cite{rakova2021responsible, sloane2022german, deng2023investigating, smith2025pragmatic, ali2023walking}. These benefits include maintaining customer trust, strengthening brand reputation, and reducing future regulatory or legal risks---all of which can ultimately minimize costs and increase overall financial gains \cite{deng2023investigating, smith2025pragmatic, ali2023walking}. Many practitioners in prior studies reported this as one of their primary strategies for convincing leadership to invest in RAI \cite{bessen2022cost, gutierrez2025explaining, schor2024mind, rakova2021responsible, madaio2020co}. Scholars from business and management backgrounds further argue that organizations cannot fully separate commercial and societal interests, and must continually balance them \cite{zimmer2022responsible, minkkinen2023co, behl2023role}---with \citet{zimmer2022responsible} (2022) 
going so far as to argue that articulating the long-term financial value of RAI should become a core element of organizational strategy. However, a number of studies \textbf{caution that over-reliance on profit-driven rationales risks reducing RAI to a branding exercise, detaching it from genuine ethical commitments}. \citet{metcalf2019owning} (2019) and \citet{mittal2024responsible} (2024) 
describe how organizations sometimes adopt fairness or transparency tools primarily for public relations purposes, without meaningfully addressing underlying harms. Similarly, \citet{hadley2024investigating} (2024) highlights cases where companies selectively implement ethical practices only when they align with business incentives, undermining their broader impact. Practitioners in many studies echoed these concerns, noting that profit-driven justifications were more likely to yield surface-level compliance than substantive ethical change. In response, \citet{ryan2024ai} (2024) and \citet{wang2024strategies} (2024) both suggest that practitioners should have accessible routes to raise concerns externally when internal channels prove insufficient through whistle-blower program 
---though how such mechanisms could be meaningfully supported by non-profits, journalists, or third-party organizations remains an open question \cite{ryan2024ai, wang2024strategies, yildirim2023investigating}.

A related but distinct motivator concerns \textbf{demand for RAI from deployers and end users}. Several studies report that practitioners' teams became more motivated to invest in RAI once customers---whether downstream deployers or end users--- demonstrated concrete interest. For example, participants in studies by \citet{gutierrez2025explaining} (2025) and \citet{schor2024mind} (2024) noted that explainability work gained organizational traction specifically when customer demand for more transparent AI systems was made visible. This framing differs meaningfully from the kinds of profit-driven arguments mentioned above: rather than appealing to long-term financial returns, generally, this strategy focuses on the immediate financial benefits of meeting specific customer demands.\looseness=-1

Finally, a number of studies highlight \textbf{compliance with existing and anticipated regulations as a powerful lever for securing organizational investment in RAI work} \cite{minkkinen2023co, orr2020attributions, ali2023walking, schiff2024emergence, ryan2023integrating, park2022designing}. Practitioners frequently invoke regulatory risk to build internal buy-in, framing RAI work not as an optional ethical exercise but as a necessary prerequisite for operating in regulated markets \cite{orr2020attributions, ali2023walking, park2022designing, nichol2023developer, elsayed2023responsible}. Crucially, both existing frameworks---such as GDPR and sector-specific regulation in finance and healthcare---and forthcoming ones---such as the EU AI Act and emerging national AI regulations---serve as signaling devices that lend RAI investment a legible, organization-wide justification extending beyond individual teams' ethical commitments \cite{schiff2024emergence, ryan2023integrating, lee2024don, bogucka2024co, hopkins2025ai}. This framing is particularly effective in organizations with established legal and compliance functions, where regulatory exposure already commands institutional attention and resources \cite{minkkinen2023co, schiff2024emergence}. We further discussed this in Section \ref{dis:regulation}.

\subsection{Lack of Tailored RAI Interventions} \label{challenge:tailored interventions}

As mentioned in Section \ref{practice:RAI-interventions}, RAI interventions such as RAI policies, processes, and tools are increasingly being adopted by RAI practitioners. However, many existing RAI interventions fall short because they are not sufficiently tailored to the specific domains, applications, and stages of the AI development lifecycle in which practitioners work. 

Across more than 100 papers in our corpus, researchers have consistently pointed to the challenge that many \textbf{RAI interventions remain too abstract to be applied to real-world AI design and development pipelines}. Early work has already shown that the RAI interventions used by practitioners are often too high-level to meaningfully inform day-to-day decisions in the AI development lifecycle \cite{passi2018trust, metcalf2019owning, holstein2019improving, orr2020attributions, madaio2020co} and called for RAI guidelines grounded in the concrete realities of model building, data curation, and system deployment. This issue, however, persists in more recent publications. For example, \citet{bughin2024doing} (2024) conducted a large-scale survey of more than 1,000 C-suite executives across the U.S. and Europe and found that organizations continue to struggle with operationalizing RAI guidelines published by major technology companies, largely because such guidelines do not map onto specific stages of the AI development pipeline. Similarly, practitioners interviewed in multiple studies published in 2024 \cite{kommiya2024towards, griffin2024ethical, kumar2024balancing, lu2024responsible} and 2025 \cite{habiba2025ml, lanne2025organisational, olson2025speaks} reported difficulty translating their companies’ internal RAI policies into actionable workflows that fit their day-to-day engineering practices. 

Beyond the challenge of being overly general, many studies have also noted that existing RAI tools and processes \textbf{disproportionately focus on evaluating model outputs, leaving other crucial stages of AI development unsupported} \cite{reuel2025responsible, rismani2023plane, kapania2023hunt, kapania2025examining, sambasivan2021everyone, sambasivan2022deskilling, ashktorab2023fairness, deng2022exploring, richardson2021towards, dominique2023factsheets, heger2022understanding}. For example, researchers have identified a lack of tools for ensuring fair and transparent data annotation workflows \cite{kapania2023hunt, kapania2025examining, sambasivan2021everyone, sambasivan2022deskilling}, despite annotation being a major source of downstream bias. Practitioners across multiple studies also reported that current fairness toolkits are primarily oriented toward evaluating model predictions, rather than supporting early-stage processes such as exploratory data analysis or feature engineering \cite{ashktorab2023fairness, deng2022exploring, richardson2021towards}. 
Likewise, many papers call for RAI policy and documentation that capture early design choices—such as problem formulation, dataset selection, and modeling assumptions—rather than only documenting model evaluation and outputs \cite{dominique2023factsheets, heger2022understanding}. \citet{reuel2025responsible} (2025) explicitly called for improving ``organizational RAI maturity'' by broadening RAI policy and documentation from isolated output level monitoring to system-level AI governance across the lifecycle of AI design and development within an organization.

Beyond stage-specific needs, practitioners across many studies highlight a substantial unmet need for \textbf{domain-specific RAI interventions}. Research examining the use of general-purpose fairness toolkits \cite{lee2021landscape}, interpretability interfaces \cite{ehsan2021expanding}, and AI ethics guidelines or checklists \cite{yildirim2023investigating, madaio2020co, bogucka2024co} consistently shows that such interventions require significant adaptation to the domain of deployment to be effective in practice. RAI concerns vary markedly across sectors. In healthcare, practitioners shared that challenges often center on clinical risk prediction, treatment prioritization, and algorithmic triage \cite{nichol2024moral, cinca2025practitioners, akbarighatar2024operationalizing}, leading practitioners to call for tools that integrate clinical guidelines and regulatory frameworks such as HIPAA. In finance, practitioners must contend with highly regulated fair lending requirements, domain-specific definitions of discrimination, and constraints on the collection and use of demographic data \cite{passi2019problem, wang2024strategies, sadek2024challenges, andrus2021we, yuan2023contextualizing}. In journalism, algorithmic curation and automated fact-checking raise distinct concerns around editorial integrity and transparency to readers, where practitioners desire support from more tailored RAI interventions that do not yet exist \cite{xiao2025might, rismani2023does}. \looseness=-1

Finally, studies of specific AI applications further underscore the need for \textbf{application-specific RAI interventions}, as responsible AI challenges vary substantially by system type \cite{smith2023many, smith2023scoping, voria2025fairness, sadek2024challenges, lanne2025organisational, yuan2023contextualizing}. For example, research on recommender systems shows that practitioners adopt fairness logics and metrics tailored to dynamic, multi-stakeholder environments, highlighting the need for tools that address exposure fairness, feedback loops, and evolving system behavior \cite{smith2023many, smith2023scoping}. Studies of conversational agents report persistent difficulties assessing harms such as toxic or hallucinated responses, emotional manipulation, and inappropriate tone using existing RAI toolkits \cite{voria2025fairness, sadek2024challenges, lanne2025organisational, yuan2023contextualizing}. Practitioners consequently call for tools that support dialog-level evaluation, conversational safety, persona consistency, and contextual appropriateness—capabilities largely absent from standard fairness libraries designed for classification tasks \cite{voria2025fairness, sadek2024challenges, kapania2025examining}. Similar needs appear across other applications, including content moderation, where practitioners seek better support for surfacing context-sensitive harms \cite{widder2022limits}; autonomous vehicles, which require scenario-based testing under uncertainty \cite{wang2022whose, voria2025fairness}; and educational technologies, which demand auditing mechanisms that account for diverse learner backgrounds and pedagogical goals \cite{rismani2023does, pant2024ethics, akbarighatar2024operationalizing}.

\subsubsection{\textbf{Opportunities for Co-designing Tailored RAI Interventions with Practitioners}}  \label{opp:co-design}

Across the literature, the majority of the papers make high-level calls for developing more pipeline-, domain-, or application-specific RAI interventions, emphasizing the limitations of one-size-fits-all approaches. However, these calls can often be abstract and stop short of offering concrete guidance.
In contrast, we identified a smaller but growing body of work that \emph{co-designs} RAI interventions in close collaboration with the potential users of these interventions, resulting in more deeply contextualized tools tailored to particular domains, organizational settings, or stages of the AI pipeline. Among the 161 papers we reviewed, 19 proposed new RAI interventions explicitly designed for particular AI applications, domains, or pipeline stages \cite{bogucka2024co, wang2024farsight, bogucka2024co, hanschke2024data, halme2024making, boyd2022designing, widder2024power, hollanek2025toolkit, bhattacharya2025explanatory, smith2025pragmatic, chen2022practitioners, lai2020perceptions, drage2024engineers, hine2024ethics, ashktorab2023fairness, van2020hiring, park2022designing, labkoff2024toward, smith2023scoping}. Many of these---especially those developed within the HCI community---were co-designed directly with industry practitioners, resulting in tools more closely aligned with real-world workflows, drawing on the unique strengths of co-design as a method to surface tacit practitioner knowledge and bridge the gap between research and practice. For example, \citet{wang2024farsight} (2024) collaborated with 10 industry AI practitioners to co-design Farsight, a tool that integrates seamlessly into early-stage prototyping workflows to support LLM harms identification. Similarly, \citet{bogucka2024co} (2024) co-designed and evaluated an AI impact assessment template with both AI developers and compliance specialists; the resulting artifact was praised for aligning with EU AI Act requirements while also being actionable within existing development processes. 
When co-designing with both technical and non-technical roles, these tools can also meaningfully bridge communication gaps, enabling faster consensus building and more actionable decision-making in RAI work across roles \cite{bogucka2024co, wang2024farsight, constantinides2024rai, elsayed2023responsible}, addressing the cross-functional collaboration challenges mentioned in Section \ref{challenge:lack of expertise}. While this emerging body of work has largely focused on stage-specific challenges, research on application- and domain-specific RAI interventions remains limited. We discuss the implications of this gap in Section \ref{dis:research}.

\subsection{\textbf{Resource and Structural Barriers to External Engagement}} \label{challenge:external friction}

Despite growing interest in involving external stakeholders in RAI efforts (see Section~\ref{practice:external}), substantial and persistent challenges hinder meaningful and sustained engagement. In particular, the challenges spanning \textbf{resource limitations, procedural and organizational barriers, and structural issues around access} collectively create friction at the interface between AI teams and external experts, end users, and annotators.

To start with, and consistent with the broader time and resource constraints discussed earlier (Section \ref{challenge:organizational dynamics}), most AI teams \textbf{do not have the resource capacities in place to effectively and meaningfully engage external domain experts or end users}. Across numerous studies, practitioners note that high-quality domain experts are expensive to hire and typically have limited availability \cite{hine2024ethics, hopkins2025ai}. For example, participants in both \citet{hine2024ethics} (2024) and \citet{hopkins2025ai} (2025) reported that attempts to involve domain experts in model evaluation failed due to mismatched timelines and insufficient budget for expert recruitment—despite strong interest from internal RAI teams to do so.\looseness=-1

As for end-user engagement, 
studies examining how teams incorporate user feedback into RAI workflows \cite{chen2022practitioners, yuan2023contextualizing, wang2023designing} describe reliance on existing UX research activities or customer-support channels. 
However, these mechanisms are rarely designed with RAI concerns in mind, resulting in limited or indirect insights into issues such as fairness, safety, or potential harms \cite{deng2023investigating, chen2022practitioners, yuan2023contextualizing}. Practitioners expressed a lack of structured pathways for incorporating end-user perspectives specifically into responsible AI processes. To this end, and in line with the opportunities outlined in Section \ref{opp:co-design}, there is a critical need for better tools and processes to incorporate user feedback throughout the AI development lifecycle, rather than only at its later stages.

Even when resources are available, \textbf{significant procedural and organizational constraints can impede collaboration with external stakeholders}. This issue is especially pronounced in highly regulated domains. For instance, all studies in our corpus focusing on responsible AI in healthcare documented developers’ difficulty in sharing internal data with clinicians or medical stakeholders due to privacy and compliance constraints \cite{nichol2023developer, nichol2024moral, labkoff2024toward, arbelaez2024integrating, bhattacharya2025explanatory}. Additionally, healthcare professionals, such as physicians and pharmacists, have limited availability for multiple design or evaluation sessions, even when developers actively attempt to involve them \cite{akbarighatar2024operationalizing, cinca2025practitioners, labkoff2024toward}. Similar challenges are surfaced for end-user engagement in RAI work. Many companies struggle to recruit participants from specific user groups---particularly marginalized communities---due to restrictions on collecting demographic data and the logistical hurdles of targeted outreach \cite{deng2023understanding, yuan2023contextualizing, kallina2025stakeholder, sarathy2023don, lee2024don}. Furthermore, leadership is often reluctant to expose internal or experimental system features to external users without clear governance processes for mitigating potential risks \cite{costanza2022audits, schiff2024emergence}. Although recent work reports growing interest in involving users in AI auditing or red-teaming efforts, practitioners consistently describe uncertainty about the appropriate procedures, approval workflows, and safeguards required \cite{deng2023understanding, wang2023designing, xiao2025might, sadek2024challenges}. 

In addition to end users and domain experts, \textbf{third-party data annotators’ participation in RAI efforts is frequently constrained by structural and organizational barriers} \cite{kazimzade2020biased, wang2022whose, sambasivan2021everyone, kapania2023hunt, sambasivan2022deskilling}. Practitioners report that annotators often lack training to proactively identify fairness or safety concerns, resulting in missed opportunities to surface RAI issues early in the annotation process \cite{wang2022whose, kapania2023hunt}. Even when issues are identified, studies consistently note limited channels for annotators to raise concerns about annotation guidelines that may encode bias \cite{kazimzade2020biased, wang2022whose, sambasivan2021everyone, kapania2023hunt, sambasivan2022deskilling}. These challenges are compounded by annotators’ working conditions—such as tight deadlines and intermediary workforce arrangements through vendors or platforms—which restrict direct engagement with AI teams and limit participation in co-design or evaluation activities \cite{wang2022whose, kapania2023hunt}. As a result, practitioners suggested that annotators’ insights that could meaningfully surface potential data-driven harms rarely feed back into RAI decision-making \cite{wang2022whose, kapania2023hunt}. More recent work further suggests that these engagement challenges are exacerbated by the use of pre-trained models, which introduce additional gaps in visibility and control over upstream data practices \cite{kapania2025examining, hedlund2025distribution}.

Finally, a small but growing body of recent work highlights \textbf{frictions that arise from the increasingly supply-chain–like structure of generative AI development}, which further complicates practitioners' ability to engage with the full range of relevant stakeholders \cite{cobbe2023understanding, balayn2025unpacking, xiao2025might, reuel2025responsible, lanne2025organisational, browne2024tech, bughin2024doing}.
Even within a single organization, AI design and development can be distributed across teams and roles, giving rise to cross-functional frictions that limit visibility and coordination \cite{subramonyam2022solving, deng2023investigating, xiao2025might}. As AI systems are assembled from components, models, and datasets produced by multiple internal teams and external vendors, responsibility becomes distributed across a diffuse ecosystem \cite{hopkins2025ai}. This distribution expands the set of stakeholders that practitioners must engage with, while each party often has limited access to other parties' resources, data, and tacit knowledge — making it difficult to address AI safety issues holistically. For example, developers working on foundation models may lack access to real-world user data held by downstream application developers, while those application developers may in turn have limited insight into the design decisions embedded in the foundation models they rely on \cite{hopkins2025ai, cobbe2023understanding}. In cases where a service provider wishes to audit their AI system for bias, it may be the application developer — in a separate organization — who holds the sensitive data necessary to do so. This fragmentation introduces what \citet{cobbe2023understanding} (2023) call an ``accountability horizon'': blurred lines of responsibility and limited visibility into upstream decisions, as documented through interviews with practitioners \cite{cobbe2023understanding}. Although supply-chain dynamics received relatively limited attention in the literature we reviewed, we expect this to become an increasingly critical challenge as the industry continues its shift toward generative AI development.

\subsubsection{\textbf{Opportunities in Developing Infrastructure for External Collaboration}} \label{opp:external}


Taken together, the challenges above point to a shared conclusion across the literature: \textbf{meaningful external engagement in RAI cannot be sustained through ad hoc efforts alone}. Instead, practitioners and scholars consistently argue for the development of more robust, well-resourced, and procedurally supported infrastructures that can lower the barriers to collaboration with end users, domain experts, and other external stakeholders \cite{hine2024ethics, chen2022practitioners, yuan2023contextualizing, wang2024operationalizing, nichol2023developer, nichol2024moral, labkoff2024toward, arbelaez2024integrating, bhattacharya2025explanatory}.

To start with, a recurring theme is the need to move beyond opportunistic or one-off engagements toward \textbf{systematic and ongoing mechanisms for participation}. Many works call for integrating domain experts and end users earlier and more continuously across the AI lifecycle---not only as sources of feedback or testers of deployed systems, but as co-creators who help shape responsible AI practices, priorities, and trade-offs from the outset \cite{bach2025insights, bar2024ai, yildirim2023investigating, sloane2022german, widder2022limits}. Achieving this vision, however, requires organizational investments in dedicated programs, roles, and workflows that explicitly support external collaboration as part of RAI work, rather than treating it as an auxiliary activity layered onto existing product or UX processes \cite{yildirim2023investigating, wang2023designing, deng2023investigating, smith2025pragmatic, madaio2020co, kallina2025stakeholder}.

Several studies suggest that one promising direction lies in the creation of \textbf{formal organizational structures designed specifically to mediate external engagement} \cite{hadley2024investigating, lanne2025organisational, madaio2020co, deng2023understanding, hopkins2025ai} . For example, inspired by Institutional Review Boards (IRBs),  \citet{hadley2024investigating} (2024) have proposed Algorithm Review Boards as a way to institutionalize cross-functional and external oversight, clarify accountability, and provide procedural guidance for engaging external stakeholders in high-stakes AI decisions \cite{hadley2024investigating}. Such boards could, in principle, offer a stable forum for incorporating perspectives from end users, domain experts, and civil society actors, while also addressing leadership concerns around risk management, confidentiality, and compliance. However, empirical studies suggest that these mechanisms remain relatively rare in practice and, where they exist, often lack clear authority, resourcing, or integration into core development workflows \cite{lanne2025organisational, hadley2024investigating}. Beyond formal review bodies, the literature also points to the importance of \textbf{supporting infrastructures that enable translation, coordination, and feedback across organizational and supply-chain boundaries}. This includes mechanisms for routing external insights, including those from user reports, annotator concerns, or expert evaluations, into decision-making processes where they can meaningfully influence the design, deployment, or governance of AI systems \cite{kapania2023hunt, deng2023understanding, pink2024trust, bogucka2024co}. Researchers acknowledge that enabling this will likely require not only new tools and methods, but also sustained organizational commitment to building infrastructures that legitimize, resource, and operationalize external collaboration as a core component of RAI practice.

\subsection{Limitations}\label{limitations} 

As mentioned in Section \ref{sec:methods}, the scope of this work is to synthesize the state of current RAI practices in industry based on the 161 empirical research we identified. This synthesis is inherently constrained by the limitations of the underlying studies we examined, including their methods, recruitment methods, studied populations, authors' positionality, and funding sources.

\section{Discussion} \label{discussion}

In the preceding two sections, we synthesized insights from the reviewed literature, many of which are directly applicable to RAI researchers, practitioners, and policymakers. In this discussion, we extend this analysis by highlighting key high-level takeaways and outlining directions for future research and practice informed by the review.

\subsection{Implications for Future RAI Research}\label{dis:research}

\subsubsection{\textbf{Understudied Areas for Future Empirical Research}}

Across the literature, researchers repeatedly surface a set of persistent challenges in RAI practice—such as limited organizational support (Section \ref{challenge:organizational dynamics}) and the abstract nature of many RAI interventions (Section \ref{challenge:tailored interventions}). While these high-level findings are now well established, we argue that future empirical research should build on this foundation by \textbf{moving beyond restating known challenges} and instead \textbf{foregrounding the \emph{contextual specificity} of RAI work} that is currently understudied. 

To begin with,\textbf{ RAI practices on \emph{generative AI} remains substantially under-studied in the empirical literature}. As of our August 2025 cutoff, the majority of studies \footnote{Except a number of works published in 2025} examine RAI practices in predictive or traditional machine learning settings, without a specific focus on generative AI. It remains unclear whether findings about documentation, accountability, expertise gaps, and organizational dynamics meaningfully carry over to foundation model development, large-scale generative systems, and increasingly complex supply chains. Given the distinctive properties of generative AI, including but not limited to the rapid capability shifts, open-ended deployment contexts, and heightened public visibility, systematic empirical engagement with RAI practice in these environments is especially critical. In line with work surfacing unique dynamics of working on AI safety in startups (Section \ref{challenge:organizational dynamics}), future work should also study the generative AI development process in ``frontier AI labs,'' such as OpenAI, Anthropic, xAI, and Google DeepMind, where the fast-paced environment and high-stakes nature of frontier model development may give rise to unique organizational dynamics, accountability structures, and other RAI challenges that differ from those observed in more established technology companies.

\looseness=-1

Future work also needs to \textbf{broaden the demographics of participants}. For example, \textbf{geographic diversity} is limited: most empirical studies focus on Western contexts, even though emerging evidence suggests that RAI priorities, regulatory pressures, and organizational dynamics vary substantially across regions. In addition, existing research often centers on a relatively narrow set of roles—such as ML engineers, RAI specialists, and UX researchers—even though many other actors materially shape RAI outcomes. Product managers, legal teams, policy staff, executive leadership, compliance officers, and procurement professionals all participate in shaping AI systems and governance processes. \textbf{Expanding the range of roles} studied would provide a more complete account of how RAI work is distributed across organizations. For example, researchers could also engage with governance practitioners---including those who translate regulatory requirements into organizational policy, negotiate compliance frameworks, or liaise with oversight bodies---in empirical study to gain more holistic views of how AI governance actually operates in practice.

Addressing these gaps will likely require greater investment in longitudinal and ethnographic approaches. Deep engagement with individual practitioners or cross-functional teams over extended periods can illuminate how RAI practices evolve, stabilize, or erode in situ—particularly in rapidly changing generative AI environments. However, access to frontline industry practitioners remains a significant constraint. Future research agendas must therefore grapple explicitly with the methodological and institutional challenges of studying high-stakes AI development settings, including questions of recruitment, transparency, and researcher–practitioner collaboration (Section~\ref{dis:practice}).

Finally, future work should also reflect critically on the practice of RAI research itself. Questions such as who RAI research primarily serves, which forms of knowledge are privileged, and how empirical findings are translated into practice warrant deeper examination. We encourage researchers to leverage our curated database on the annotated papers to conduct meta-analyses of the current RAI research landscape, identify structural blind spots, and chart future research directions for the field. In parallel, we are already conducting complementary meta-analyses using this dataset in a separate paper.

\subsubsection{\textbf{Toward empirically grounded and effective RAI interventions}}

Given that many challenges in RAI practice have been consistently documented over time, we argue that the field is \textbf{well positioned to move from primarily cataloging problems toward building and field testing RAI interventions} that can withstand real-world constraints. To this end, we encourage more research that explicitly designs RAI interventions informed by empirical findings and evaluates how these interventions function in practice. As mentioned in Section \ref{opp:co-design}, recent work that engages in co-design with practitioners demonstrates the promise of this approach. Building on these efforts, future research can further leverage participatory and co-design methods to develop interventions tailored to specific domains, applications, and lifecycle stages (Section~\ref{challenge:tailored interventions}). In addition, given the recurring challenges around collaboration and coordination across roles (Section~\ref{challenge}), we call for research that prioritizes the design of boundary objects—such as reports, narratives, templates, and visualizations—that facilitate shared understanding, reduce misinterpretation, and support communication across organizational and disciplinary boundaries.

The rise of generative AI further underscores the need for this shift. When developing RAI interventions, future empirical studies should take into account the full generative AI supply chain, including foundation model providers, vendors, downstream developers, and deployers. Examining these multi-actor ecosystems can help researchers understand how responsibility, documentation, accountability, and trust are distributed---and often fragmented---across organizational boundaries. Developing methods to trace these dynamics across the supply chain represents an important direction for future work.

Finally, rather than treating RAI interventions as isolated metrics, checklists, or dashboards, researchers should \textbf{conceptualize RAI as an end-to-end sociotechnical system} encompassing training, incentives, workflows, decision rights, escalation pathways, and organizational accountability structures situated in real world industry settings (Section \ref{challenge:tailored interventions}). Studying RAI in this holistic manner better reflects how RAI work is actually carried out in practice, and can also support better RAI practice on the ground. This also implies a need to empirically investigate how different interventions function in concert, rather than evaluating each mechanism in isolation. In addition, future work should move beyond one-off interviews or workshops toward longitudinal and ethnographic approaches that enable deeper understanding of how RAI practices evolve over time. Such approaches—including sustained field engagement, interview and survey saturation, and follow-up studies that examine how previously identified insights translate into practice—can provide stronger empirical grounding for understanding organizational dynamics and the durability of RAI interventions. While this type of research is often challenging due to access, recruitment, and resource constraints, as suggested in previous section, prior work has demonstrated its feasibility in select cases \cite{passi2019problem}. In some contexts, auto-ethnographic or embedded research—where researchers become closely integrated into organizational workflows—may offer a viable pathway for capturing situated RAI practice while maintaining reflective distance.

\subsection{Implications for Future RAI Practice} \label{dis:practice}

Taken together, our synthesis of both reported ``success stories'' (Section~\ref{practice}) and persistent challenges (Section~\ref{challenge}) suggests that effective RAI practices depend less on any single tool or intervention or actor, and more on whether organizations cultivate the conditions for RAI to be sustained over time. These conditions include practitioners’ baseline awareness and competence around RAI, alignment between individual incentives and organizational priorities, the integration of RAI interventions into real-world development workflows, and supported interfaces with external stakeholders. Building on the opportunities identified across Section~\ref{challenge}, we highlight two broader implications for future RAI practice.

\subsubsection{\textbf{Build RAI Capability as Organizational Infrastructure}}

First, organizations should treat \textbf{RAI capability as core organizational infrastructure}, rather than as an ad hoc or auxiliary activity. As discussed in Section~\ref{challenge:lack of expertise}, many practitioners report limited access to RAI expertise and training. Instead of framing RAI education as a one-time onboarding requirement or optional resource, organizations should invest in ongoing, cross-role learning that evolves alongside changing technical and regulatory landscapes. This includes recurring training tailored to different roles (e.g., engineers, product managers, designers, policy specialists), opportunities for shared sensemaking across functions, and mechanisms for updating guidance as new risks and evaluation challenges—particularly those introduced by generative AI—emerge.

Beyond training, \textbf{RAI capabilities must be embedded into everyday work practices}. As shown in Sections~\ref{challenge:organizational dynamics} and \ref{challenge:tailored interventions}, RAI efforts are more durable when they are integrated into existing development processes—such as product reviews, experiment tracking, model evaluations, and launch gates—rather than positioned as parallel or optional work streams that are easily deprioritized under time and budget pressure. As many paper suggested as the current challenges and implications (Section \ref{challenge:tailored interventions} and \ref{challenge:organizational dynamics}), one failure mode for doing RAI is to only doing RAI work at the very end before deployment, or even after deployment. We therefore call for organizations to integrate RAI considerations directly into core delivery pipelines, ensuring that responsible AI work is treated as part of ``how work gets done'' rather than as an additional layer of compliance.

Finally, sustaining RAI practice requires clear organizational ownership and resourcing. Organizations should \textbf{establish explicit responsibility for RAI outcomes, including defined roles, dedicated staffing, and escalation pathways when concerns arise}. Without such structures, RAI work risks becoming under-resourced, dependent on individual goodwill, or reduced to performative documentation. Explicit decision authority and accountability mechanisms can help ensure that RAI considerations meaningfully influence product and organizational decisions. However, as mentioned throughout our findings (Section \ref{challenge}), small-size startups and rapidly scaling companies often operate under distinct constraints related to resources, timelines, and governance structures. How to design sustainable RAI ownership and accountability mechanisms under such startup constraints remains an open question for both researchers, policymakers, and practitioners to systematically address.

\subsubsection{\textbf{Institutionalize External Engagements}}

Second, organizations must more deliberately \textbf{institutionalize external engagement} as a core component of RAI practice. As generative AI systems are increasingly assembled from third-party models, datasets, and services, accountability becomes more diffuse—particularly for external experts, end users, and data annotators whose input must traverse organizational and contractual boundaries to effect change. Meaningful external participation cannot rely on goodwill or informal outreach alone; it requires procedural support that makes participation safe, legitimate, and actionable even under privacy constraints and vendorized development pipelines.

To address these frictions, companies could establish formal structures such as vetted expert panels, algorithm review boards, structured red-teaming programs, and protected channels for annotators to surface concerns. Practically, this involves budgeting for participation, developing compliant data-sharing procedures, and designing clear pathways for external insights to inform internal decision-making. Emerging cases where third-party organizations have been engaged by frontier AI labs---such as OpenAI and Anthropic---in pre-deployment risk evaluation \cite{hurst2024gpt, dziemian2026} offer early models for operationalizing this kind of external engagement. However, as discussed in Section~\ref{challenge:organizational dynamics}, the for-profit nature of these companies may complicate the integrity of such procedures.

Yet even well-structured external engagement carries inherent tensions. When companies fund or commission external evaluators, the resulting work is not truly independent—researchers collaborate on assessments the company has chosen to conduct, on terms it controls. A related gap is the field's limited capacity to learn from failures: unlike safety-critical industries with structured incident review processes, AI development lacks robust mechanisms for surfacing and learning from high-profile breakdowns.  Addressing these limitations will require progress on two complementary fronts: making RAI research more accessible and actionable for practitioners (requiring effort from both researchers and practitioners), and enabling collective learning across organizations through structured information-sharing, even among competitors.

\subsection{Implications for Existing and Future AI Regulation}
\label{dis:regulation}

Our synthesis suggests that AI regulation is most likely to be effective when it is designed around how RAI work is actually performed and coordinated within and across organizations, rather than around idealized or purely normative models of compliance. Throughout Section \ref{challenge}, a recurring concern across the literature is that existing and proposed AI-related regulations are often perceived by practitioners as overly abstract, overly ambitious, or difficult to operationalize in practice. To address these challenges, future regulation should prioritize \textbf{implementability}, ensuring that regulatory requirements map onto concrete lifecycle stages, decision points, and organizational workflows. 
When requirements align with how AI systems are actually developed, evaluated, and deployed, practitioners are more likely to translate regulatory expectations into sustained practice rather than treating them as high-level principles disconnected from day-to-day work \cite{selbst2021institutional}.

Another important implication from synthesizing prior empirical work is that regulation should focus \textbf{not only on regulating artifacts (e.g., models, datasets, or documentation), but also on supporting the buildout of organizational capacity}.
Given persistent gaps in expertise, training, and institutional support documented throughout the literature, governance frameworks should incentivize or require the development of RAI capabilities, such as dedicated roles, ongoing training programs, internal review structures, and clear escalation pathways. Without attention to capacity, regulatory compliance risks devolving into symbolic documentation that does little to improve real-world outcomes.

Relatedly, future AI regulation should \textbf{prioritize substantive accountability over procedural compliance to ensure meaningful change rather than ``symbolic compliance''}---the formal adoption of governance structures, documentation, or review processes that satisfy regulatory expectations without materially altering underlying practices or risk outcomes \cite{edelman2020working, bamberger2015privacy, waldman2021industry}. In the papers we reviewed, practitioners frequently critique regulatory approaches that increase compliance burden without corresponding impact (Section \ref{challenge:tailored interventions}). 
Many organizations are formalizing RAI roles, committees, documentation processes, and internal review procedures (Section \ref{practice:professionalized}); yet such professionalization can devolve into symbolic compliance if it does not meaningfully shape product and deployment decisions. This dynamic is well documented beyond AI governance. Legal scholarship shows how ambiguous regulatory mandates can incentivize firms to institutionalize procedural safeguards that courts treat as evidence of due diligence, even when underlying practices remain unchanged \cite{edelman2020working, waldman2021industry}. Edelman’s work on the professionalization of HR in response to anti-discrimination law, for example, demonstrates how compliance structures can become decoupled from substantive reform \cite{edelman2020working}, a pattern later observed in the privacy domain by Waldman \cite{waldman2021industry}. Our review suggests that similar dynamics may already be emerging in RAI practice. To mitigate this risk, regulation should focus less on exhaustive reporting requirements and more on embedding accountability within organizational decision-making authority. Frameworks that require organizations to justify trade-offs, document mitigation rationales, and demonstrate how identified risks are acted upon may be more effective than rigid templates or one-time “check-the-box” assessments.

Finally, many practitioners report struggling to keep pace with rapidly evolving model capabilities and failure modes (Section \ref{challenge:lack of expertise} and \ref{challenge:organizational dynamics}). Even prior to the current generative AI surge, organizations faced gaps in expertise, shifting technical baselines, and uncertainty about how to operationalize high-level governance principles. Importantly, many papers reviewed indicate that\textbf{ the \textit{anticipation} of regulation can shape organizational behavior even when enforcement is not yet in place}. The prospect of regulatory scrutiny often motivates companies to formalize internal RAI policies, allocate resources, professionalize RAI roles, and invest in internal governance infrastructure (Sections~\ref{practice:professionalized} and \ref{challenge:organizational dynamics}). From this perspective, regulation functions not only as an enforcement mechanism but also as a signaling device that influences organizational priorities and capacity-building. Future regulators could strategically leverage this signaling function to shape companies' RAI practices—but only if \textbf{regulation is itself designed to evolve alongside the technology it governs}. Static requirements risk becoming quickly outdated; adaptability must therefore be embedded in regulatory architecture from the outset, through strategies such as technologically neutral statutory language, delegated interpretive authority, and frameworks with explicitly revisable components. Recent examples illustrate this approach: the EU AI Act enables iterative refinement through updateable annexes and delegated acts \cite{EUAIAct}, while the NIST AI RMF playbook provides domain-specific implementation guidance and encourages organizations to share experiences through case studies and shared repositories \cite{nist2023aimf}. By building adaptability into regulation's structure, policymakers are better positioned to respond to emerging risks, capability shifts, and unintended consequences as generative AI systems continue to advance.



\bibliographystyle{ACM-Reference-Format}
\bibliography{citation, industryRAIpractice}

\appendix

\section{Appendix} \label{appendix}

\subsection{Keyword search}

\textit{[[Abstract: practices] OR [Abstract: perceptions]OR [Abstract: needs] OR [Abstract: challenges]] 
AND [[Abstract: empirical] OR [Abstract: qualitative] OR [Abstract: quantitative]OR [Abstract: interview] OR [Abstract: survey]] 
AND [[Abstract: industry] OR [Abstract: company]] 
AND [[Abstract: "responsible ai"] OR [Abstract: "rai"] OR [Abstract: "fairness"] OR [Abstract: "harm"] OR [Abstract: "ethics"] OR [Abstract: "bias"]]  OR [[Abstract: "safe"] OR [Abstract: "safety"] OR [Abstract: "alignment"] OR [Abstract: "red-team"] OR [Abstract: "red team"]] 
AND [[Abstract: "ai"] OR [Abstract: "artificial intelligence"] OR [Abstract: "machine learning"] OR [Abstract: "ml"] OR [Abstract: "algorithm"] OR [Abstract: "decision-making"] OR [Abstract: "decision making"] OR [Abstract: "autonomous"]]}

\end{document}